\documentclass[a4paper,11pt,natbib]{article}

\usepackage{pos}
\usepackage{natbib}

\usepackage{slashed}

\title{The Physics of Neutrinoless Double Beta Decay: A Primer}

\author*[a]{Ben Jones}
\affiliation[a]{University of Texas at Arlington\\Department of Physics, 108 Science Hall, 502 Yates St, Arlington, TX 76019}
\abstract{Neutrinoless double beta decay is a hypothetical radioactive process which, if observed, would prove the neutrino to be a Majorana fermion: a particle that is its own antiparticle. In this lecture mini-series I discuss the physics of Majorana fermions and the connection between the nature of neutrino mass and neutrinoless double beta decay.  We review Dirac and Majorana spinors, discuss methods of distinguishing between Majorana and Dirac fermions, and derive in outline the connection between neutrino mass and double beta decay rates.  We conclude by briefly summarizing the experimental landscape and the challenges associated with searches for this elusive process.}

\FullConference{%
  Theoretical Advanced Study Institute: The Obscure Universe: Neutrinos and Other Dark Matters - TASI2020\\
  1-26 June, 2020\\
  Boulder, Colorado, USA
}
\tableofcontents
\begin{document}
\maketitle

\section{Introduction}

This document is a writeup of a two-lecture mini-course given at the
TASI2020 Summer School. Previous speakers had given excellent lectures
on the mechanisms of neutrino mass generation and neutrino oscillation
experiments. My goal was to convey something essential and intuitive
about the physics of neutrinoless double beta decay ($0\nu\beta\beta$)
- the most sensitive known way to test for the Majorana nature of
the neutrino. 

The day-to-day work of researchers in the field of neutrinoless double
beta decay searches primarily revolves around controlling experimental
backgrounds to the mind-bendingly and vanishingly low levels required
for sensitivity. If observed, $0\nu\beta\beta$ would be the slowest
process ever observed. The experimentally allowed half-life of the
process is now more than $t_{1/2}\geq10^{26}$ years in all practical
isotopes. Some practitioners contribute to this quest through development
of new technologies, others develop new analysis methods, and still
others develop new radio-pure materials, or devote their
careers to advancing the precision of radio-assays. The field is a
sprawling and exciting one, and especially for instrumentation enthusiasts,
$0\nu\beta\beta$ remains a topic that rewards creativity: both the need for and possibility
of a technological breakthrough are real, and a Nobel-class discovery
with existential implications may be just around the corner.

In these lectures we
will discuss the physics of $0\nu\beta\beta$
in a way that is hopefully comprehensible for graduate students who
have taken a particle physics class from one of the classic
textbooks, for example, Thomson~\cite{thomson2013modern} or Halzen and Martin~\cite{halzen2008quark}.  I will spend almost
 no time reviewing the many kinds of beautiful and complicated
experiments that search for this process, though good concise reviews exist elsewhere~\cite{gomez2011sense,dolinski2019neutrinoless,elliott2002double}. I will also sweep vast numbers of important theoretical details under
the rug, in the interest of conveying the underlying principles. The tone will be conversational and informal throughout, though for those desiring a more formal treatment, many are available~\cite{Haxton:1984ggj,bilenky2012neutrinoless,bilenky2015neutrinoless,engel2017status}.  My intention is that the explorations presented here will be interesting, convey an intuitive picture of the underlying physics, and potentially function as a springboard for students first familiarizing themselves with the subject to use to dive into more complex and specialized literature.
\\
\\
Without further ado, let us get the party started.

\section{Majorana and Dirac Spinors}

We begin our
explorations with a brief review of the spinors that represent Dirac
and Majorana neutrinos. This will help motivate what is to come.

\subsection{Dirac spinors in the Weyl basis}

Neutrinos are relativistic, massive spin 1/2 particles. As such, they obey
the Dirac equation, whose solutions are necessarily four-component
objects called spinors~\cite{peskin2018introduction,thomson2013modern}, $\psi$:
\begin{equation}
\left(i\gamma^{\mu}\partial_{\mu}-m\right)\psi=0\label{eq:Dirac-1}.
\end{equation}
Depending on what formalism we are using, the Dirac equation can be interpreted either a Schrodinger-like equation for a four-component wave function (that is, probability densities for finding four distinct kinds of particle in that place - as in relativistic quantum mechanics) or as an equation of motion for a four-component field (that is, an object representing a finite amount of ``particle-ness'' at every place for four distinct kinds of excitation, as in quantum field theory).  In either case, the four entries in $\psi(x)$ carry four (or technically ``up to four'', as we will soon see) pieces of information about what exists at each point in space. What each of those four quantities  independently represent is not trivially stated, and depends on the basis in which $\psi$ is written.

With a judicious choice of basis, the four-component Dirac spinor
can be written in terms of two upper components representing a left
chiral field $\xi$, and two lower components representing a right
chiral field $\eta$:
\begin{equation}
\psi=\left(\begin{array}{c}
\xi\\
\eta
\end{array}\right)\label{eq:Weyl}.
\end{equation}
This basis is called the Weyl basis, and in it the $\gamma$ matrices
have the form:
\begin{equation}
\gamma^{\mu}=\left(\begin{array}{cc}
0 & \sigma^{\mu}\\
\bar{\sigma}^{\mu} & 0
\end{array}\right),\quad\quad\gamma^{5}=\left(\begin{array}{cc}
-I & 0\\
0 & I
\end{array}\right).
\end{equation}
We can verify that the top two components are left chiral and the
bottom two are right chiral by appealing to the chirality operator
$\gamma^{5}$. Left chiral spinors are eigenstates of $\gamma^{5}$  with eigenvalue -1, and right-chiral spinors or eigenstates of $\gamma^{5}$ with eigenvalue +1. Considering the Weyl basis spinors, we find:
\begin{eqnarray}
\gamma^{5}\left(\begin{array}{c}
\xi\\
0
\end{array}\right)=\left(\begin{array}{c}
-\xi\\
0
\end{array}\right),\quad\rightarrow\quad\mathrm{left\,chiral}
\\
\gamma^{5}\left(\begin{array}{c}
0\\
\eta
\end{array}\right)=\left(\begin{array}{c}
0\\
\eta
\end{array}\right),\quad\rightarrow\quad\mathrm{right\,chiral}
\end{eqnarray}
This is a special property of the Weyl basis. Other ways of writing
the Dirac equation in terms of different $\gamma$'s and different
$\psi's$ can be obtained by $\psi\rightarrow U\psi$ and $\gamma=U\gamma U^{-1}$
leading to spinors with other useful properties of spinors, but this convenient upstairs-downstairs decomposition by chirality is special for the Weyl basis. 

Looking at the matrix structure of the $\gamma's$ we see mass term
in the Dirac equation is block diagonal in this basis, whereas the
kinetic term is not; thus in general the two chiral parts will get
mixed up with each other as time evolves. Writing out the Dirac equation
in terms of Eq. \ref{eq:Weyl}:
\begin{eqnarray}
i\left(\partial_{0}+\vec{\sigma}.\nabla\right)\eta=m\xi\label{eq:Weyl1},
\\
i\left(\partial_{0}-\vec{\sigma}.\nabla\right)\xi=m\eta\label{eq:Weyl2}.
\end{eqnarray}
The two sub-fields are coupled; this means you might start out with
something left chiral, but apply a little time evolution and it will
become a somewhat right chiral - chirality is not conserved. The exception
is when $m=0$; then the two sub-fields decouple from one another
and evolve independently. 
\begin{eqnarray}
\partial_{0}\eta=-\vec{\sigma}.\nabla\eta\label{eq:Weyl1-1},
\\
\partial_{0}\xi=\vec{\sigma}.\nabla\xi\label{eq:Weyl2-1}.
\end{eqnarray}
For massless particles, it is thus possible to consider forever-left-chiral and
forever-right-chiral solutions to the Dirac equation, with chirality
conserved. One way to understand this is that as $m\rightarrow0$,
chirality becomes equivalent to helicity, and helicity of a free particle is always conserved
due to the rotational symmetry of the Universe.

\subsection{Introducing the Majorana fermion}

Lets get one thing straight: whether the neutrino is a Majorana or
a Dirac spinor, it can surely be represented by a four component object
that looks like $\psi$ and evolves according to the Dirac equation.

\medskip{}

So what do we mean when we talk about a possibly non-Dirac nature of the
neutrino spinor? It is not a question about whether the spinor evolves
according to the Dirac equation, it always does. The question
is instead: ``are $\xi$ and $\eta$ independent fields that
can have independent values everywhere in space?''. Or conversely,
``if you know what $\eta(x)$ is at some place, can you determine
what $\xi(x)$ is there from it, or do you need more information?'' 

If given $\eta(x)$ you can know $\xi(x)$, you have a Majorana
spinor. If given $\eta(x)$ you cannot know $\xi(x)$ without more information, you have a
Dirac spinor. That is the entirety of the distinction, though we will soon see why it leads to interesting consequences. A Dirac spinor has four entries so four degrees of freedom at every point in space
- we can interpret these degrees of freedom as independent
left- and right-handed fermions and anti-fermions. A Majorana spinor
has only two degrees of freedom at every point in space, since if
you know two of the components (say, the two components of $\eta)$, you immediately
know the other two (the two components of $\xi$), so they are not degrees of freedom. 

Before we can ask deeper questions, like ``what do the two degrees of freedom of the Majorana spinor represent?'', we should start by asking: can
such a solution to the Dirac equation as we have described exist? Are we smart
enough to construct a spinor such that not only does $\eta=f(\xi)$
at some initial time, but this property is maintained forever as the
field is driven forward in time by the Dirac equation? Ettore Majorana
was smart enough to come up with one, when he wrote down this spinor:
\begin{equation}
\psi=\left(\begin{array}{c}
\xi\\
-i\sigma_{2}\xi^{*}
\end{array}\right)\label{eq:MajoranaSpinor}.
\end{equation}

This object is neat because it both satisfies the Dirac equation and
the Majorana condition. Whatever the top two components are at a given
time, the bottom two will stay related to them as in Eq. \ref{eq:MajoranaSpinor}.
Of course, because it obeys Dirac's equation it
evolves in a way that is consistent with special relativity; satisfies
the Einstein energy momentum relationship $E^{2}=p^{2}+m^{2}$ (i.e.
it represents a particle of mass $m$); and has spin 1/2. 

Since only two of the components of the four-component field are independent,
and the Dirac equation involves four distinct time evolution rules,
two of them must be unnecessary for the Majorana spinor. Indeed, for this object the four-component
Dirac equation is four-component overkill, because an equally good
two-component equation of motion can be obtained by substituting Eq.
\ref{eq:MajoranaSpinor} into Eq. \ref{eq:Weyl1}:
\begin{equation}
\sigma^{\mu}\partial_{\mu}\xi+m\sigma_{2}\xi^{*}=0.
\end{equation}
This is called the Majorana equation of motion. The Dirac equation
is still good if we find it useful, but this one is just as good,
and has less components. One property of the full four-component spinor
that looks quite interesting is this one:
\begin{equation}
\psi^{c}\equiv i\gamma^{2}\psi^{*}=\psi\label{eq:ChargeConj}.
\end{equation}
Which you can prove just by applying the charge conjugation operation
to Eq. \ref{eq:MajoranaSpinor}. Recall that this mathematical manipulation
is exactly what you would do to turn a particle spinor into an antiparticle
one, or vice versa. The Majorana spinor has the intriguing property
that the charge conjugate of the spinor is the spinor itself; the particle
is its own antiparticle. This is in contrast to the Dirac spinor with
four components, where $\psi^{c}=i\gamma^{2}\psi^{*}\neq\psi$, and
the particle and antiparticle are distinct. 

So there we have it, a Majorana spinor has two degrees of freedom
which correspond to "amount of left handed thing at each point in space" and "amount of right handed
thing at every point in space", and both of those things are constructed such that they are both particle and antiparticle at the same time.

\subsection{What kinds of particles can be Majorana fermions?}

Not every kind of particle can be a Majorana fermion. To begin with,
a Majorana particle needs to have a mass (otherwise the distinction between
Majorana and Dirac fermions becomes irrelevant), and have spin 1/2
(otherwise it wouldn't satisfy the Dirac equation). There are other
constraints, too. Consider a particle satisfying the Dirac equation
but interacting with gauge fields, for example, a particle interacting
electromagnetically with the photon field $A$. The equation of motion
for $\psi$ would be:
\begin{equation}
\left(i\gamma^{\mu}\left(\partial_{\mu}+iqA_{\mu}\right)-m\right)\psi=0.\label{eq:Dirac}
\end{equation}
We can take the complex conjugate and multiply by $i\gamma^{2}$ to
find an equation of motion for $\psi^{c}$:
\begin{equation}
\left(i\gamma^{\mu}\left(\partial_{\mu}-iqA_{\mu}\right)-m\right)\psi^{c}=0.
\end{equation}
This is notably different to Eq. \ref{eq:Dirac}. If have a field
where $\psi=\psi^{c}$, the above two equations would be in contradiction
to one another. Except in the special case where $q=0$, when they
would be consistent. If we wish to have a field where $\psi=\psi^{c}$
both now and at all subsequent times as governed by the Dirac equation
of motion, the charge of the fermion must be zero. The same argument
applies to all the gauge charges, not only the electric charge, which
would impose similar consistency constraints. Thus the only fermions
that can satisfy the Majorana condition as the field evolves are ones that
carry no gauge charges. We only know one such fermion in the standard
model: the neutrino.

So the neutrino we know and love might be Majorana fermion, whereas
no other known particle can. But is it? There are several compelling
reasons to suspect that perhaps it is:
\begin{enumerate}
\item A neutral particle with a small Majorana mass is the first hint one
would expect to observe from new high-scale physics, were the standard
model a low energy effective theory. This is because the relevant
term in the Lagrangian that generates Majorana neutrino masses - the
``Weinberg operator''~\cite{weinberg1979baryon} - is the only dimension 5 operator (a term
suppressed by only one power of some new high energy scale rather than
more powers) consistent with the gauge symmetries of the standard
model. This is a bit outside our scope today, but worth knowing.
\item Experimentally we have only observed two kinds of neutrino
- the left handed thing we usually call a neutrino and the right handed
thing we usually call an antineutrino. It would be economical if these
were the only two components of the field, not two of four, with other
two being mysteriously unobservable.
\item Majorana neutrinos are a low energy prediction of leptogenesis~\cite{fukugita1986barygenesis}, a
compelling theoretical mechanism for explaining the matter/antimatter
asymmetry of the contemporary Universe. This asymmetry is vital for
our existence, but is presently unexplained by the known laws of physics.
If neutrinos are both Majorana and CP-violating, they may have played
a key role in generating this asymmetry.
\end{enumerate}
That all feels strongly suggestive that Majorana nature is a property of the neutrino that is well worth testing for.  

\section{Neutrinoless double beta decay as a short baseline neutrino experiment}
Majorana fermions are well motivated, and the neutrino may very
well be one. Whether it is or not is a question that must be addressed
with data. We now turn our attention to what data we should seek.

\subsection{A thought experiment: searching for Majorana neutrinos in a neutrino beam}

The question for the experimentalist is how to test whether
the neutrino is indeed its own antiparticle. Naively this sounds
like it shouldn't be too difficult. In neutrino oscillation experiments
we make and study neutrinos and antineutrinos all the time, after
all. How can we check whether they are the same or different?

\begin{figure}
\begin{centering}
\includegraphics[width=0.95\columnwidth]{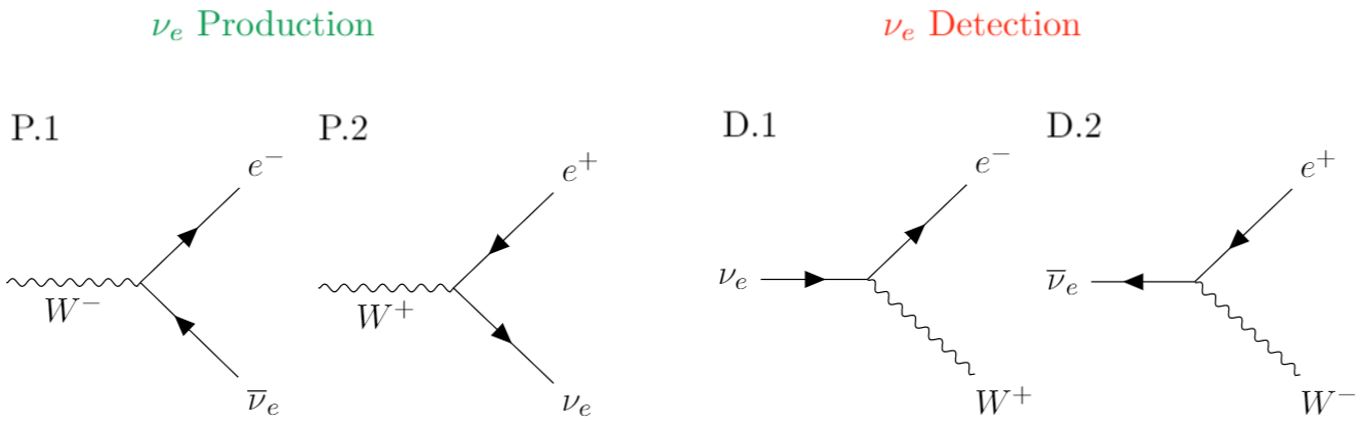}
\par\end{centering}
\caption{The allowed interactions of electron neutrinos and antineutrinos.
\label{fig:The-allowed-interactions}}

\end{figure}

First let us ask, how do we tell whether we're studying neutrinos
or antineutrinos in these experiments? An obvious difference between
neutrinos and antineutrinos is that the neutrinos produce negative
leptons in weak charged current scattering interactions, whereas the
antineutrinos produce positive leptons. In our naive picture, without
Majorana fermions, this would be considered to be a necessary consequence
of lepton number (L) conservation - one lepton goes in (L=1) and one
lepton goes out (L=1) or one anti-lepton (L=-1) goes in and one anti-lepton
goes out (L=-1). The allowed production and detection modes of $\nu_{e}$
and $\bar{\nu}_{e}$ with their accompanying charged leptons, assuming
neutrinos and antineutrinos are distinct and lepton number is conserved,
are shown in Fig. \ref{fig:The-allowed-interactions}.

\begin{figure}
Neutrino discovery channel by Project Poltergeist
\begin{centering}
\includegraphics[width=0.8\columnwidth]{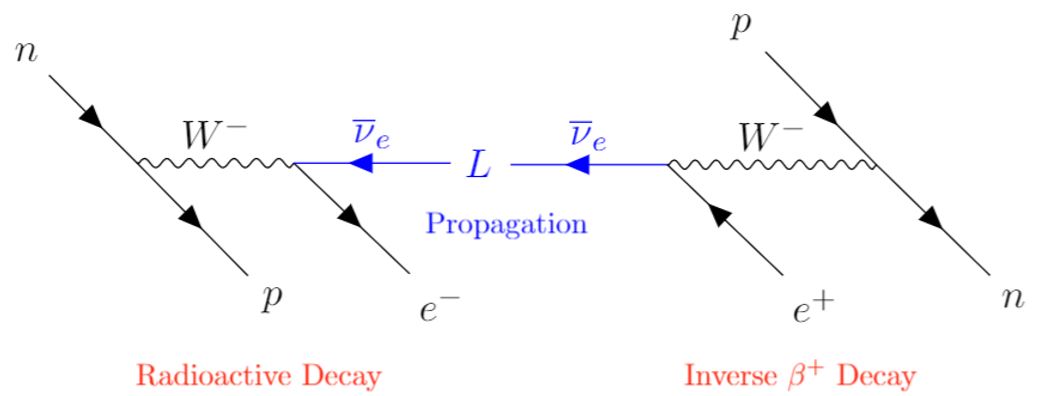}
\par\end{centering}
\begin{centering}
Possible detection channel if neutrinos are their own antiparticles?
\par\end{centering}
\begin{centering}
\includegraphics[width=0.8\columnwidth]{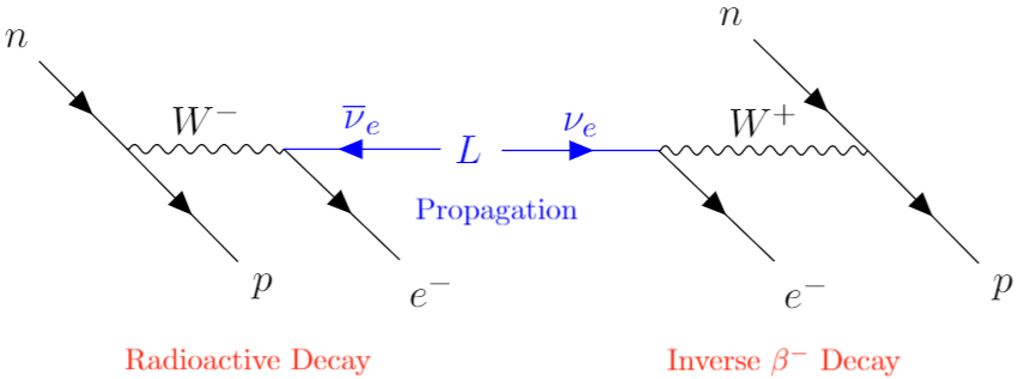}
\par\end{centering}
\caption{Top: The process by which the neutrino was first detected by Project
Poltergeist - electron neutrino production by beta decay followed
by electron neutrino detection by inverse $\beta^{+}$ decay. Bottom:
A similar process that we might imagine could be allowed if neutrinos
are their own antiparticles. \label{fig:Top:-The-way}. }
\end{figure}

The discovery of anti-neutrinos (seen long before neutrinos) was first
made using the process P.1 of Fig. \ref{fig:The-allowed-interactions} to produce $\bar{\nu}_{e}$ from nuclear reactors, and process
D.2 for $\bar{\nu}_{e}$ detection. The schematic outline of the nobel-prize
winning 1955 experiment, called ``Project Poltergeist''~\cite{reines1953detection}, is shown
in Fig. \ref{fig:Top:-The-way}, top. Electron antineutrinos are produced
in $\beta^{-}$ decays within a nuclear reactor. They propagate to
a liquid scintillator, where some small fraction undergo ``inverse
beta decay'', creating a positron and a free neutron. The positron
scintillates immediately and the neutron bounces around and eventually
captures producing a delayed, detectable signature. This double
pulse signature in liquid scintillator is characteristic of low
energy antineutrino detection. Seeing such events with rates
correlated with reactor activity led to the conclusive discovery of
the anti-neutrino.

If the neutrino is equivalent to the anti-neutrino then in principle
we might imagine another process, where we first produce a $\bar{\nu}_{e}$
alongside an $e^{-}$ (we look at the electron so we know it was produced
as an antineutrino), but we see it interact as if it were a $\nu_{e}$,
producing another $e^{-}$ (we look at the electron so we know it
interacted as a neutrino). Clearly such a process would violate lepton
number conservation, since we had no leptons (L=0) in the beginning
and we have two electrons (L=2) the end. This lepton number violation
is the hallmark of Majorana neutrinos. Observing this hypothetical
process, shown in Fig. \ref{fig:Top:-The-way}, bottom, would establish
the neutrino as a Majorana fermion using a neutrino beam, and earn us an all expenses paid
ticket to Stockholm.

When we delve a little deeper, we find that this experiment that looks
so easy, is in fact, not so. It is thwarted by the competing forces
of the chiral structure of weak interactions and angular momentum
conservation. Recall that weak interactions have a vector-minus-axial-vector
(V-A) structure, which means that the fermion currents in these Feynman
amplitudes have a $\gamma^{\mu}\left[1-\gamma^{5}\right]$ sandwiched
between two fermion fields. Mathematically the current at the first
neutrino interaction vertex looks like:
\begin{equation}
\bar{u}_{e}\frac{1}{2}\gamma^{\mu}\left[1-\gamma^{5}\right]v_{\nu}=\bar{u}_{e}\gamma^{\mu}P_{L}v_{\nu}.
\end{equation}
On the other hand, the one at the second neutrino interaction vertex
looks like:
\begin{equation}
\bar{v}_{e}\frac{1}{2}\gamma^{\mu}\left[1-\gamma^{5}\right]u_{\nu}=\bar{v}_{e}\gamma^{\mu}P_{L}u_{\nu}.
\end{equation}
Recall also that most confusing of factoids from that class you once
took on particle physics: $P_{L}$ selects the left chiral parts of
particle spinors and the right chiral parts of anti-spinors (Thomson~\cite{thomson2013modern}
P140). Thus the outgoing neutrino-like thing at the left vertex must
be left-chiral, and the incoming antineutrino-like thing at the right
vertex must be right chiral, per the utterly screwy\footnote{Pun intended.}
preferences of the weak interaction. 

The chiralities of the emitted and detected things don't match, so
we would be dead in the water with this experiment if
chirality were a conserved quantity. Luckily we have seen that chirality is not conserved
for massive fermions, so maybe it's ok. What might be a problem though,
is that the neutrino's social mobility is severely limited by considerations
of helicity.

Recall that in the ultra-relativistic limit, when $m\ll E$, helicity
and chirality are equivalent. In this limit, the emerging ``antineutrino-like
thing'' at the left vertex would have to be right-helicity, which
means having spin pointing left-to-right in this picture; whereas
the entering ``neutrino-like thing'' at the right vertex would have
to be left-helicity, meaning spin pointing right-to-left. This change
of spin mid-flight, assuming the neutrino doesn't interact with anything
on the way, is prohibited by angular momentum conservation. And, even if
we have Majorana neutrinos we still have angular momentum conservation.
So for a massless neutrino this process would surely be prohibited.

Only in the case where the neutrino is somewhat non-relativistic and
chirality and helicity are inequivalent can this process occur without
violating spin conservation\footnote{``Being somewhat non-relativistic'' is of course an unnecessarily
fancy way of saying ``having a non-zero mass''. }.  In this case, the neutrino leaving the left vertex (left chiral) has
a small but non-zero right helicity component. Explicitly, for high
energy particles the decomposition of chiral states $u_{L,R}$ in
terms of helicity $u_{\uparrow,\downarrow}$ states takes the form, for example for u$_L$:
\begin{equation}
u_{L}=\frac{1}{2}\left(1+\kappa\right)u_{\uparrow}+\frac{1}{2}\left(1-\kappa\right)u_{\downarrow},\quad\quad\kappa\sim\left(1-\frac{m}{E}\right).
\end{equation}
Neutrinos are very light, so always very relativistic, and this implies
their $\kappa$ is very close to 1. The tiny $u_{\downarrow}$ component,
proportional to $m_{\nu}/E_{\nu}$, is what can allow the process
we have sketched out to proceed. We can use this information to estimate
the rate of this process relative to the ``boring'' lepton-number-conserving one that was
used to discover the antineutrino. Rates of processes
are proportional to the squares of their amplitudes, by Fermi's Golden
Rule. Our hypothesized lepton-number-violating ($\Delta L=2$) process
will thus be suppressed relative to the Standard Model lepton-number-conserving
($\Delta L=0$) process by a factor $m_{\nu}^{2}/E^{2}$. 

We don't know the neutrino mass, but we do know for sure that it is
less than around 1eV~\cite{aker2019improved}. If we consider reactor neutrino experiments~\cite{wen2017reactor}, the beams
have energy of a few MeV. Thus, at most one in every $\sim2\times10^{13}$
neutrinos that interact at all, might interact according to our hypothetical
lepton-number-violating mode. The rest will interact via the good-old-fashioned,
salt-of-the-Earth lepton number conserving mode, even if neutrinos
are indeed Majorana fermions. 

And here lies the rub: unfortunately, probing processes that are suppressed
by factors of more than $10^{12}$ remains far outside the statistical
reach of even the most barn-burning reactor neutrino experiments,
which collect around 400 interactions per day. Accelerator and atmospheric
neutrino experiments can't help either - their event rates are also far from sufficient, and the energies are even higher, given an even large rate suppression for the L=2 process. Worse still, even if we could
detect the $10^{13}$ neutrino interactions needed to have a high
probability of observing this new process one time, the prospect of
suppressing backgrounds from the standard model process to the part
per trillion level sounds basically insane, far beyond any existing
experimental capabilities. 

Alas, our thought experiment is probably not a viable way to detect
Majorana neutrinos. But perhaps it can serve as a guidepost toward
a more viable approach.

\subsection{The ultimate short baseline experiment}

\begin{figure}
\begin{centering}
\includegraphics[width=0.49\columnwidth]{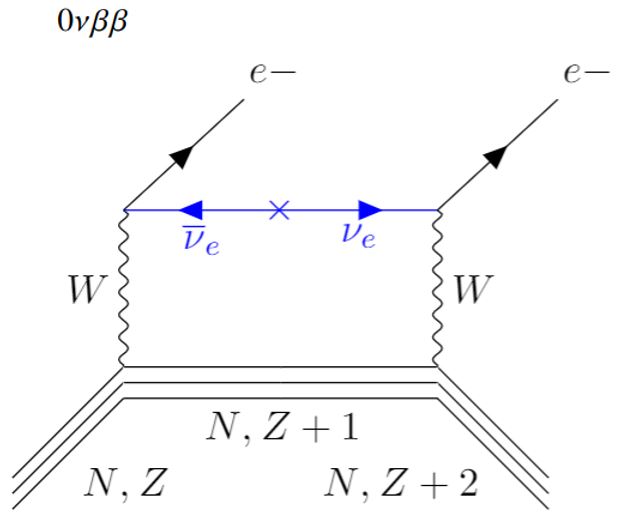}\includegraphics[width=0.49\columnwidth]{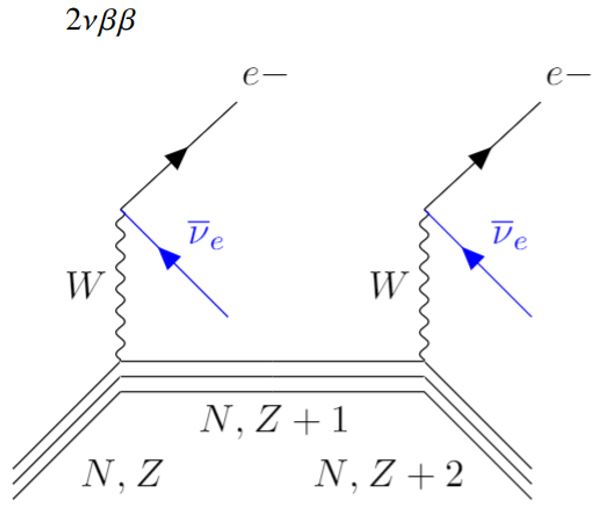}
\par\end{centering}
\caption{The Feynman diagram for neutrino-less double beta decay (left) and
two neutrino double beta decay (right) \label{fig:The-Feynman-diagram}}

\end{figure}

Consider the process shown in (Fig.~\ref{fig:The-Feynman-diagram}).
In this process, a nucleus emits a $W$ boson, turning a neutron into
a proton. This creates an electron and an antineutrino; this neutrino
propagates a very short distance (on the order of the size of a nucleus),
and if Majorana, interacts again with the same nucleus. The ultimate
result is production of two electrons, while turning two neutrons
into two protons. This process is the exact analog of the hypothetical
lepton number violating process we discussed above, but with a baseline
smaller than the size of a nucleus. This process is called neutrino-less
double beta decay, $0\nu\beta\beta$ for short, and searching for
it is the most compelling known way to test for the Majorana nature
of the neutrino.

It is worth taking a moment to muse on the question, why might this
process be any different from the one we discussed before, in terms
of plausibility of detection? Are we not again simply overwhelmed
with background from a similar $\Delta L$=0 process with the ``right-sign''  leptons,
and $e^{+}e^{-}$ in the final state, as before? The answer can be
found by considering what happens to the nucleus in this situation.
We would, in that case, need to turn a proton to a neutron and back
again. Assuming we start in the ground state of the nucleus as in
ordinary matter, our only option would be to change one
nucleus into a similar looking one, but with one more proton one and one
less neutron, and then flip this one back to exactly where we started.
No other final state is energetically accessible from the ground state.
In this process there would be no free energy to put into the electrons, apart from a tiny quantity of order a few eV absorbed by the nuclear recoil.
For the process to go, we would need at least enough energy difference
available between the initial and final nuclear states to make two
electron masses, however - with zero available energy, the two new electrons simply
cannot be created. 

On the other hand, the $\Delta L$=2 process where two
neutrons change into two protons actually transforms the nucleus
into a different one. If we pick a nucleus where a heavier isotope
would turn into a lighter one when two neutrons turned into two protons,
we have a setup that seems almost custom-made for driving forward
the lepton-number-violating $\Delta L=2$ process while suppressing
its lepton-number-conserving $\Delta L=0$ counterpart. In such a system, the energy
released can be converted into electron mass energy and kinetic energy in the two-electron final state.

\subsection{The pairing force as the engine of double beta decay}

The effect that makes searching for this weird nuclear decay a viable
possibility rather than science fiction is
the nuclear pairing force. Nucleons in nuclei have spin, and so they
have magnetic moments. Just like electrons in atomic orbitals, it
is usually energetically favorable for them to pair up with
spins opposing each-other in equivalent spatial orbitals.  This maximizes
wave-function overlap and the stabilizing effect of the attractive
spin-opposite spin interaction. The result is that nuclei with even
numbers of protons, or even numbers of neutrons, are nearly always
slightly more tightly bound than similar nuclei with odd numbers of
both. In then yet more special even-even nuclei where both proton and neutron numbers are even, all of the nucleons can
pair up in this harmonious, stabilizing way.

\begin{figure}[t]
\begin{centering}
\includegraphics[width=0.49\columnwidth]{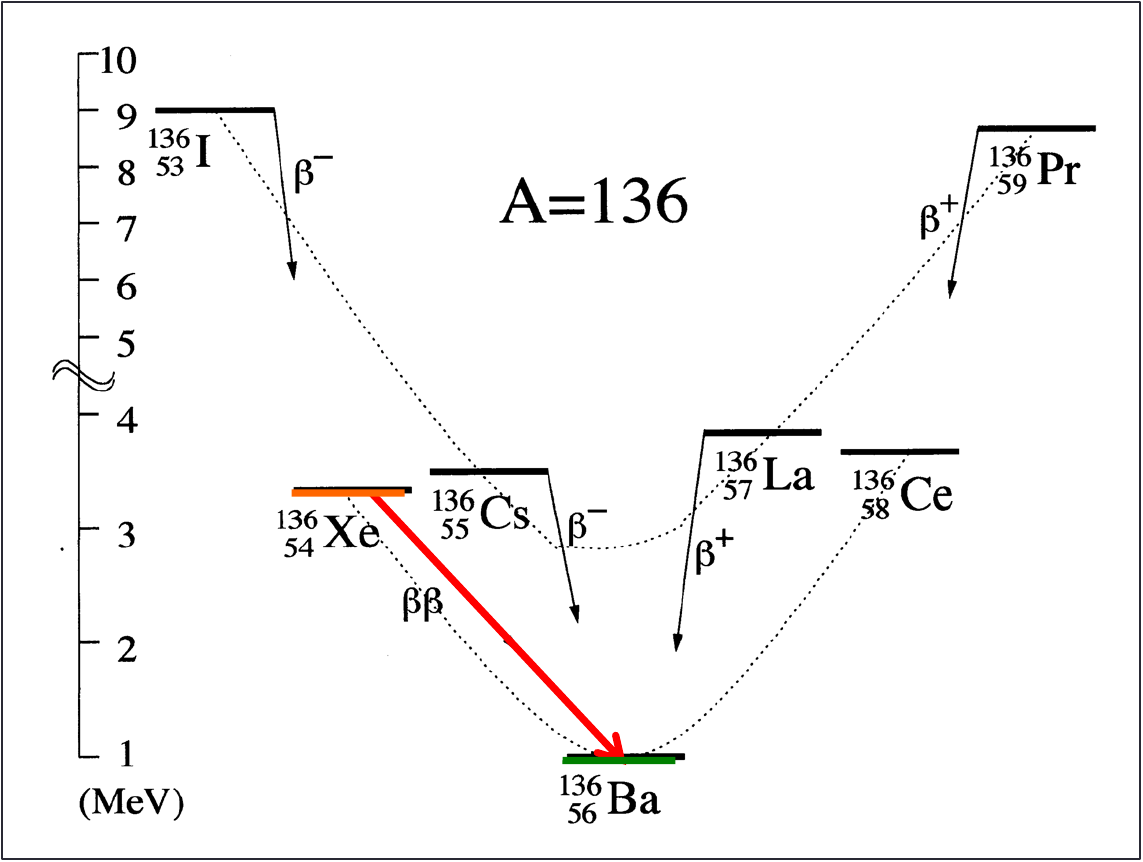}$\quad$\includegraphics[width=0.4\columnwidth]{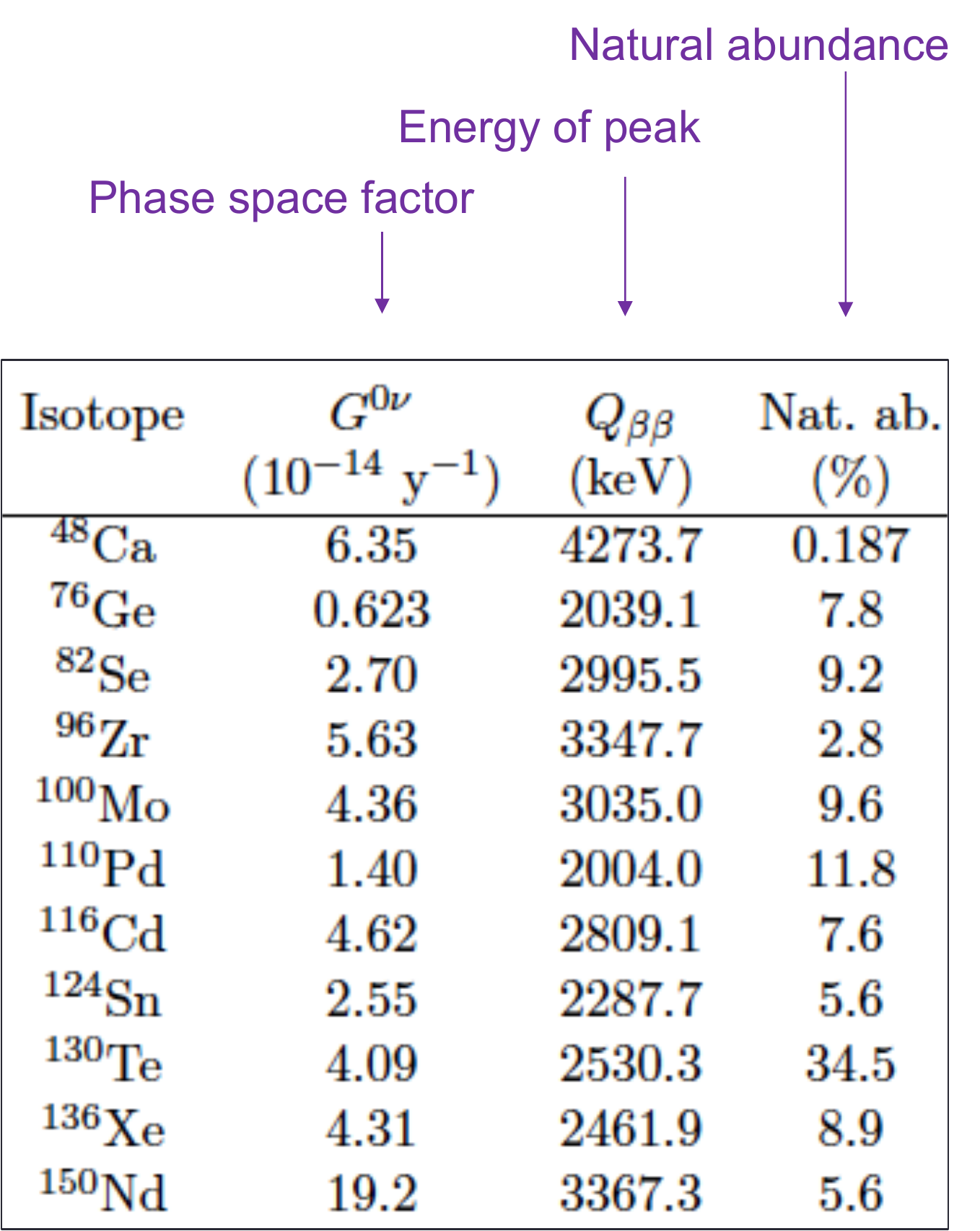}
\par\end{centering}
\caption{Left: Masses of isotopes in the N=136 isobar. Right: table of phase space factors, Q-values, and natural abundances of double beta decaying nuclei.\label{fig:Pairing}}
\end{figure}
Beta decays, of either the single or double variety, never change
the total number of nucleons in a nucleus (this would violate baryon
number, which is never violated perturbatively in the standard model).
So beta decays always move us within an ``isobar'', that is, the
collection of nuclei with the same total number of nucleons N but
different numbers of protons Z.  When we plot the mass of the nucleus vs the number of protons within
an isobar, we immediately see the effect of the nuclear pairing force.
Figure \ref{fig:Pairing}, left shows an example. Even-even
nuclei $^{136}$Xe and $^{136}$Ba are stabilized by pairing. Odd-odd
nuclei $^{136}$I and $^{136}$Cs are less tightly bound. The result
is that $^{136}$Xe is energetically forbidden from undergoing single
beta decay to $^{136}$Cs, but is energetically allowed to double
beta decay to $^{136}$Ba. This makes $^{136}$Xe an
example of an outstanding nucleus we might be able to use as a laboratory
to search for $0\nu\beta\beta$. There are a handful of such nuclei
where the nuclear pairing force makes the energetics work just-so,
some examples given in Fig.~\ref{fig:Pairing}, right.
So maybe the Nobel-worthy experiment we dreamed up in the last section
can work after all, if we  make our neutrino baseline smaller than the size of the
nucleus.

Not so fast - just like the earlier
thought experiment, of course, the rate of $0\nu\beta\beta$ is still
suppressed by a factor of $m^{2}/E^{2}$. This is because the neutrino
still has to interact with the ``wrong'' helicity at the second
vertex. Naturally if $m_\nu=0$, the neutrino would have no ``wrong helicity''
component and so the rate of $0\nu\beta\beta$ would be zero. The
rate would also be zero if the neutrino mass is finite but the neutrino
is not Majorana. Only if the neutrino is both massive and Majorana,
can $0\nu\beta\beta$ proceed with a non-zero rate, and the suppression
by $m^{2}/E^{2}$ means that the rate of this decay is going to be
slow as heck. Since we don't yet know the neutrino mass we don't
know what decay rate is expected, but existing constraints imply a
decay half-life to the neutrinoless mode of at least $\geq10^{26}$
years in most practical isotopes.

\subsection{Neutrinoless and two neutrino double beta decay}

Although we don't suffer from the ``non-helicity-flipping'' background
of our earlier thought experiment, there is still a decay process of even-even
isotopes that does represent a challenging background to $0\nu\beta\beta$
. This is two neutrino double beta decay, an event where two ordinary
beta decays happen at once, with two electrons and two neutrinos in
the final state. The $2\nu\beta\beta$ mode goes at least $10^{6}$
times faster (in xenon) than the predicted rate of the $0\nu\beta\beta$
mode. Diagrams of this processes are shown in Fig.~\ref{fig:The-Feynman-diagram},
right. 

What are we going to do about this background? Well, the final state
of this two-neutrino process is rather different from the process we are interested in. It has two
electrons and two neutrinos, whereas the $0\nu\beta\beta$ has only
two electrons. The question is whether we can we tell the difference
reliably enough to reject backgrounds with this different final state.
Answering this question positively is the first criterion for a sensitive
$0\nu\beta\beta$ experiment.

Naively we might imagine telling the difference between the two modes
by detecting the neutrinos and vetoing the events that have them.
However, given the interaction cross sections of neutrinos at this
energy, to reliably tag an emerging neutrino would require a block
of detector material of around a light year in length - not a very
viable prospect. It seems we are restricted to only measuring the
charged decay products: the electrons and possibly the daughter nucleus.

But there is still a difference between the final states of $0\nu\beta\beta$ and $2\nu\beta\beta$, even if we can only measure the electrons. In
the case of the two neutrino mode, the available energy is shared
between two detectable electrons and two unobservable neutrinos. In
the case of the neutrino-less mode, all the energy is directed into
the electrons. Since the amount of energy available is fixed - it
is the mass difference between the parent and daughter nuclei - the
neutrino-less mode should produce a mono-energetic spike, whereas
the two-neutrino mode produces a broad spectrum\footnote{As an aside, we are reminded that the difference between an energy
spike when all the decay products are observed, and a smear where
some escape undetected, is what first suggested the existence of the
neutrino to Pauli in 1928~\cite{pauli1991earlier}.}.

\begin{figure}
\begin{centering}
\includegraphics[width=0.49\columnwidth]{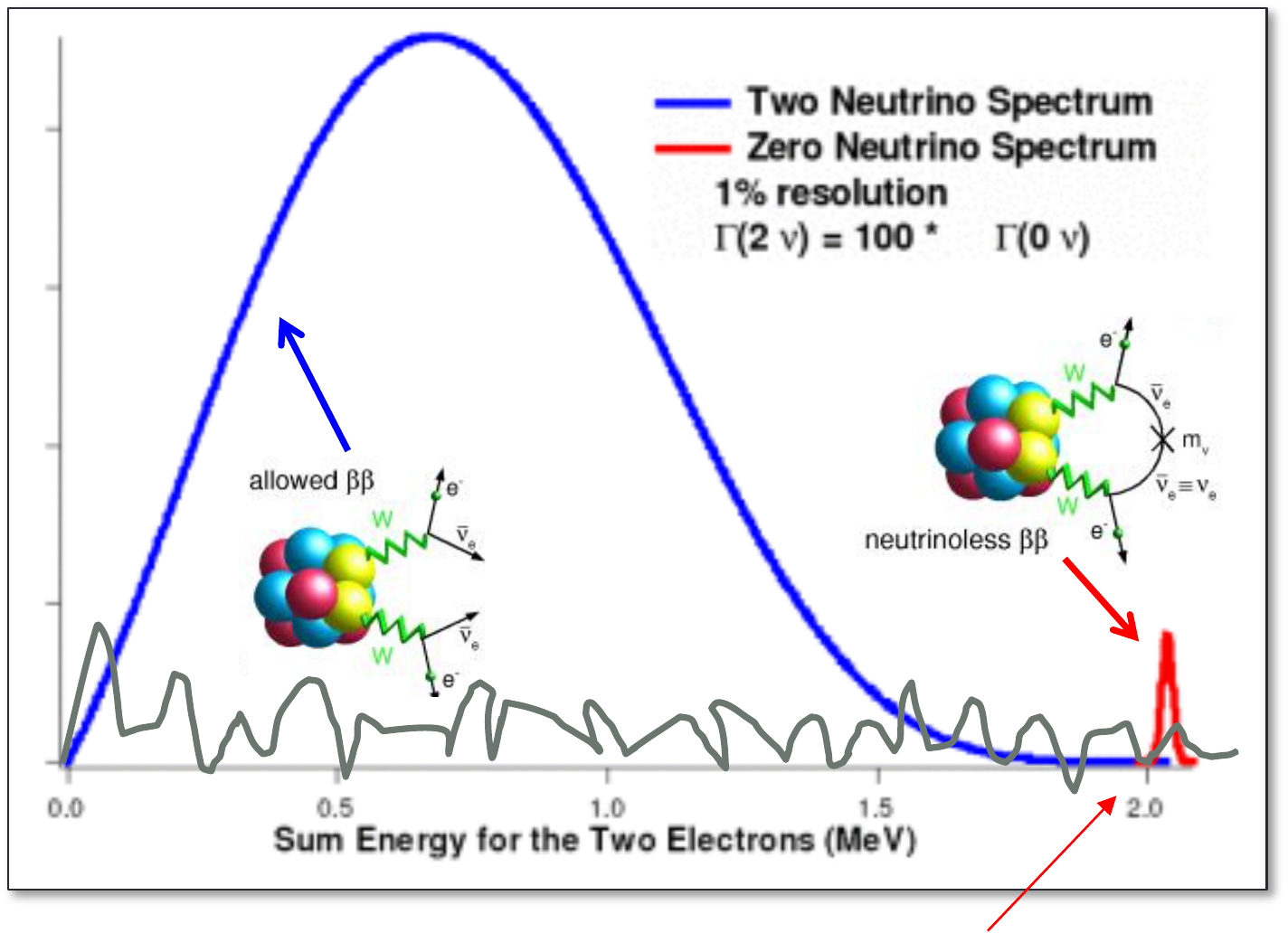}\includegraphics[width=0.49\columnwidth]{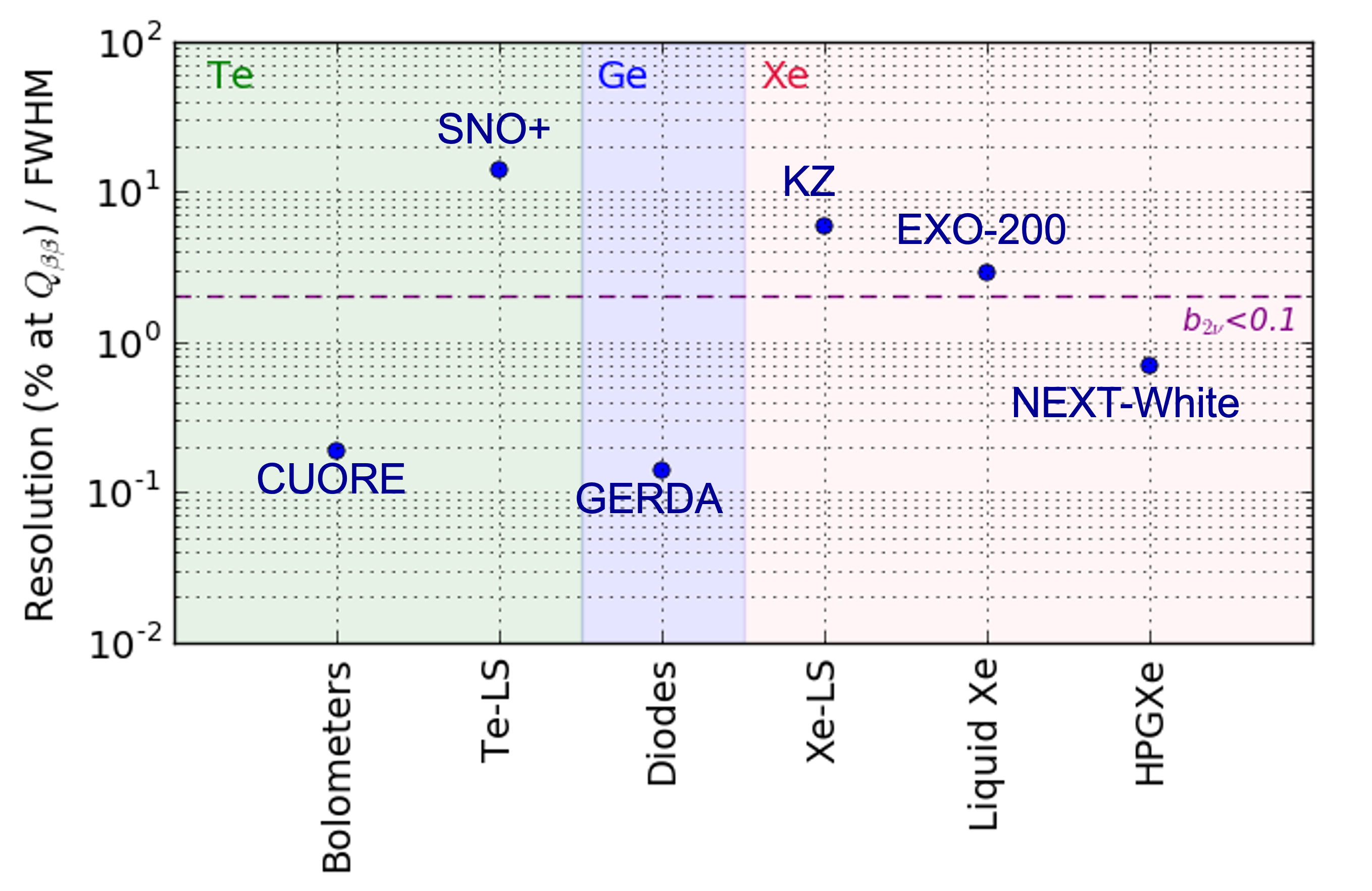}
\par\end{centering}
\caption{Left: Spectrum of energy emitted into electrons in the two-neutrino
and neutrino-less modes of double beta decay. \label{fig:Cartoon-showing-energies}.
Right: demonstrated energy resolutions in various detector technologies, tellurium bolometers, tellurium in liquid scintillator, germanium diodes, xenon in liquid scintillator, high pressre xenon gas.}
\end{figure}

While the energy is in general shared between final state particles,
its sharing is inherently random. It is possible to have almost all
the energy in the electrons and almost none in the neutrinos. Thus
there is a tail to the $2\nu\beta\beta$ spectrum that runs almost
(up to two neutrino masses) all the way up to the hypothetical $0\nu\beta\beta$
peak. In order not to incorporate background from this tail, exquisite
energy resolution in the electron energy measurement is required. 

In order to
to reduce the background from $2\nu\beta\beta$ in the $0\nu\beta\beta$
region of interest to a suitable level, generally agreed by the community
to be around 0.1 counts per ton per year, next-generation $0\nu\beta\beta$ experiments require energy resolution of at most
most 2\% FWHM. More precise resolutions protect from
the effects of inevitable non-Gaussian tails in energy distributions.
The energy resolutions demonstrated in various technologies to date are shown in Fig. \ref{fig:Cartoon-showing-energies},
right. Several of them comfortably meet this goal; others still suffer
from $2\nu\beta\beta$ backgrounds in their energy regions of interest.
Progress toward ever longer lifetime sensitivities will require this background
to be ever more  efficiently mitigated via precise energy resolution, if it is not to become an irreducible barrier to sensitivity.

\section{The rate of $0\nu\beta\beta$}

Calculating the rate of either neutrinoless or two-neutrino double
beta decay is a tricky business, because it involves understanding
the properties of the nucleus that is undergoing decay. Here we present
a valiant attempt that illustrates a lot of the key physics, but necessarily
involves skipping steps and making simplifications that are not made
in modern theoretical calculations. I regret nothing.  Let us start with the Hamiltonian, following Refs.~\cite{Haxton:1984ggj,fukugita2013physics}:
\begin{equation}
H=\left(\sqrt{2}G_{F}\left|V_{ud}\right|\right)^{2}\left[\bar{e}_{L}\gamma_{\mu}(1-\gamma_{5})\nu_{L}\right]\left[-\bar{\nu}_{L}^{c}\gamma_{\mu}(1-\gamma_{5})e_{L}^{c}\right]\left[\bar{p}\gamma^{\mu}(1-g_{A}\gamma_{5})n\right]\left[\bar{p}\gamma^{\mu}(1-g_{A}\gamma_{5})n\right]\label{eq:Hamiltonian}
\end{equation}
Many elements of this object are familiar; we have the Fermi constant
twice - expected for a second order weak process; the CKM matrix element
$|V_{ud}|$, since the $W$ bosons here couple u quarks to d quarks; Two terms
for leptonic vertices, though one of them is a bit weird looking compared
to what we might expect in a more garden-variety process due to those
conjugate operations - we will be contracting this with the other
term to form a propagator soon; and then two hadronic currents that
turn protons into neutrons. We also see the weak vertex terms: $\gamma^{\mu}(1-\gamma_{5})$
as expected at the leptonic vertices, but $\gamma^{\mu}(1-g_{A}\gamma_{5})$
at the hadronic ones. It is worth taking a brief digression to explain
this, since the value of $g_{A}$ is an ongoing topic of discussion
you might hear about, in connection with theoretical predictions of
the rate of double beta decay theory.

\subsection{The role of vector / axial vector currents and $g_{A}$}

Why in the vertex for the nucleon did the vector part have a coupling
strength of 1, as in the leptonic sector, but the axial part have
a coupling strength of $g_{A}$? Before asking why the axial coefficient
is modified but the vector one is not, we can ask a more pressing
question - why would any of these coefficients not be 1? The weak
charged current interaction couples like $\gamma^{\mu}(1-\gamma_{5})$,
right? What is this $g_{A}$?

\begin{figure}
\begin{centering}
\includegraphics[width=0.49\columnwidth]{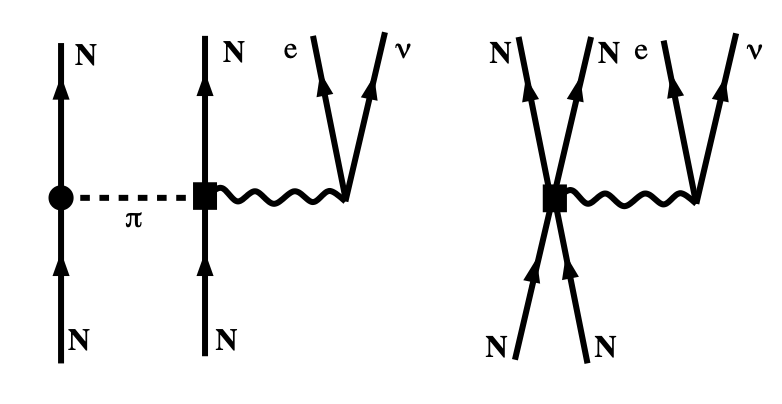}
\par\end{centering}
\caption{Meson exchange currents that ``renormalize'' $g_{A}$ in beta decay.
\label{fig:Meson-exchange-currents}}
\end{figure}

The answer is that nuclear effects, such as meson exchange currents
in the nuclear medium, affect the strength of the couplings that drive
beta decay within the nucleus. This can make life miserable
for everyone, even talented nuclear theorists. Some examples of these
processes are shown in Fig. \ref{fig:Meson-exchange-currents}. The
consequence is that the neutrino never really interacts with just
one nucleon in a vacuum, it interacts with a nucleon that is mid-interaction
with all the other nucleons all the time, and so the charges that
couple to the weak interaction are not just 1 as they would be for
single, fundamental particle interactions.

The next question is why the axial charge seems to be modified in
the above expression, but not the vector charge. The reason a mysterious
thing called the Conservation of Vector Current (CVC) hypothesis.
It is an idea greatly pre-dates the standard model itself, but seems to be largely robust even
with all the new effects the full standard model of weak interactions
piles on top of the old assumptions about nucleons. 

Consider the proton and neutron as a doublet in isospin, such that:
\begin{equation}
\Psi=\left(\begin{array}{c}
p\\
n
\end{array}\right).
\end{equation}
Under this construction, these two particles are simply manifestations
of a simpler, more general particle, the nucleon, differing only in
that they have a different position in the isospin doublet. People
used to think about them that way, and it turns out that to a good
degree of approximation, they behave so, especially in scenarios where
the strong interaction is the dominant force. The 3-component of isospin
$T_{3}$ of the proton and neutron are determined by applying $\tau_{3}$,
which is just the Pauli matrix $\sigma_{3}$ with a fancy name, to
make sure we remember it acts in isospin space and not in spin space:
\begin{eqnarray}
\tau_{3}p=\tau_{3}\left(\begin{array}{c}
1\\
0
\end{array}\right)=+\left(\begin{array}{c}
1\\
0
\end{array}\right),\quad T_{3}(p)=+1.
\\
\tau_{3}n=\tau_{3}\left(\begin{array}{c}
0\\
1
\end{array}\right)=-\left(\begin{array}{c}
0\\
1
\end{array}\right),\quad T_{3}(n)=-1.
\end{eqnarray}
The electromagnetic current that couples photons to nucleons only
notices the proton, since the neutron is neutral. Thus we can write
the coupling to the photon field $A_{\mu}$ as:
\begin{equation}
\left[\bar{p}\gamma^{\mu}p\right]A_{\mu}=\left[1\times\bar{p}\gamma^{\mu}p+0\times\bar{n}\gamma^{\mu}n\right]A_{\mu}=J_{EM}^{\mu}A_{\mu}.
\end{equation}
In terms of isospin projectors, the electromagnetic current is given
by:
\begin{eqnarray}
J_{EM}^{\mu}=\bar{\Psi}\left(1+\tau_{3}\right)\gamma^{\mu}\Psi,\\
=\bar{\Psi}\gamma^{\mu}\Psi+\bar{\Psi}\gamma^{\mu}\tau_{3}\Psi,\\
=g_{I}J_{iso-scalar}+g_{V}J_{3}.\label{eq:EMb}
\end{eqnarray}

Where in the second equality the current is broken up into an isoscalar
part, and a part that depends on the third isospin matrix $\tau_{3}$,
and we injected a $g_{I}=1$ and a $g_{V}=1$. The second term can
be thought of as the third element in a 3-vector in isospin space,
involving $\left(\tau_{1},\tau_{2},\tau_{3}\right)$. 

Were isospin really an exact symmetry, that is, protons and neutrons
are essentially the same except for their difference in isospin, then
isospin would itself be a conserved quantity.
\begin{equation}
\partial_{\mu}J^{\mu}=\partial_{\mu}\bar{\Psi}\vec{\tau}\gamma^{\mu}\Psi=0\label{eq:Isospin}.
\end{equation}
That this is nearly true is in fact why such a weird thing as isospin
was ever useful to anyone; all the nucleons in all the nuclei are
sufficiently similar to one another from the point of view of the
strong force that this conservation law appears true in a great number
of scenarios, especially those whose dynamics are dictated dominantly
by strong interactions; in others it is broken by a little bit. But
renormalization of the nuclear charges, what we are talking about
right now, is governed by strong interactions.

Eq. \ref{eq:Isospin} is in fact not one conservation law but three,
one for each element in $\vec{\tau}$. Thats one way to look at the
situation, anyway. Another is to say, this is one conservation law
but of a vector quantity $\vec{\tau}$. To the extent that isospin
is a good symmetry, rotating the direction of all the $\vec{\tau}$'s
of all the particles by some fixed angle would lead to an equally
viable universe that would obey the same laws as our own, albeit starting
from a different initial condition. It is hard to imagine such a universe;
but were the strong interaction calling all the shots this weird thought
experiment would work out just fine.

What does all this have to do with renormalization of weak processes?
Well, intriguingly, the currents that couple to the W boson in double
beta decay involve isospin raising and lowering operators constructed
in terms of other elements of that same isospin vector, just different
components:

\begin{eqnarray}
J_{Weak}^{\mu}=g_{V}\bar{\Psi}\left(\tau_{1}\pm i\tau_{2}\right)\gamma^{\mu}\Psi-g_{A}\bar{\Psi}\left(\tau_{1}\pm i\tau_{2}\right)\gamma^{5}\gamma^{\mu}\Psi
\\
=g_{V}J_{Weak,V}^{\mu}-g_{A}J_{Weak,A}^{\mu}\label{eq:Weakg}
\end{eqnarray}
Which has been split into vector and axial vector parts, $J_{Weak,V}^{\mu}=J_{1,V}\pm iJ_{2,V}$ and $J_{Weak,A}^{\mu}=J_{1,A}\pm iJ_{2,A}$ Even though they couple to different gauge bosons than the electromagnetic
current, and generate apparently different phenomena, the vector parts
they invoke are just like the electromagnetic currents, but with different
elements of the $\vec{\tau}$ vector. No matter what weird QCD stuff
happens, higher order nuclear effects cannot modify the charge of
nucleon, since whatever stuff those effects might violate they surely
conserve charge. Thus the electromagnetic current, Eq. \ref{eq:EMb}
is unmodified by nuclear effects. If isospin symmetry is really a
symmetry, then, the $g_{V}$ in Eq. \ref{eq:EMb} is also unmodified
by higher order nuclear effects - it is driven by the same kind of
currents, just with a different choice of direction for the isospin
vector $\vec{\tau}$. This is the consequence of CVC hypothesis in
$0\nu\beta\beta$ - to the extent that isopsin is a good symmetry,
$g_{V}=1$ even with nuclear effects. 

On the other hand, all bets are off concerning $g_{A}$. This quantity
is not related to anything to do with electromagnetism and is not
protected by the principle of CVC. Generally one expects it to be
modified by potentially large amounts, due to higher order nuclear
effects. Since $g_{A}$ will feature to the fourth power in the rate
of $0\nu\beta\beta$, even small changes to it matter
a lot. 

The theoretical treatments are evolving. For the present moment, $g_{A}$
is a significant source of uncertainty in the nuclear currents involved
in double beta decay. Measurements from neutron decay (a decay which
also involves both vector and axial vector parts) presently give the
most precise value, $g_{A}\sim1.27$. However, it is debatable whether
this can be taken as an accurate value for the complex nuclei involved
in neutrinoless double beta decay experiments. Two-neutrino double
beta decay rates suggest that $g_{A}$ could be significantly modified
relative to neutron decay. The most modern matrix element methods
claim a self-consistent calculation which incorporates the effects
of $g_{A}$ fully. It seems reasonable to be hopeful for a conclusive
resolution of the question renormalization of $g_{A}$ question in
the relatively near future. 

\subsection{The leptonic and hadronic tensors in single light neutrino exchange}

The Hamiltonian Eq. \ref{eq:Hamiltonian} can be taken as being a
reasonable one for both two-neutrino and neutrinoless decays. In the
case of neutrinoless double beta decay, the neutrino that is emitted
at one vertex is absorbed at the other one. In quantum field theory
this corresponds to replacing the two $\nu$ fields in the above expression
with a fermionic propagator, and it turns out that the propagator
here is the same one we would expect for Dirac fermions~\cite{fukugita2013physics}:
\begin{equation}
\frac{\slashed k+m}{k^{2}-m^{2}}.
\end{equation}
The matrix element for the process then takes the form:
\begin{eqnarray}
{\cal M}=-2G_{F}^{2}\left|V_{ud}\right|^{2}\sum_{n}\int\frac{d^{4}k}{\left(2\pi\right)^{4}}\bar{u}(p_{1})\frac{1-\gamma_{5}}{2}\gamma_{\mu}\frac{\slashed k_{\nu}+m}{k_{\nu}^{2}-m^{2}}\gamma_{\nu}\frac{1-\gamma_{5}}{2}v(p_{2})e^{i\vec{k}_{\nu}.r}\times\\
\langle N_{i}|J^{\mu}|n\rangle\langle n|J^{\nu}|N_{f}\rangle2\pi\delta(k_{\nu}-E_{1i}+E_{1n}+\epsilon_{1}).
\end{eqnarray}
Where $|n\rangle$ is the intermediate state of the nucleus. This
can be written in terms of a hadronic and a leptonic part:
\begin{equation}
=-2G_{F}^{2}\left|V_{ud}\right|^{2}\sum_{n}L_{\mu\nu}^{n}H_{n}^{\mu\nu}.
\end{equation}
The leptonic part has a form that looks like it might be manageable:
\begin{equation}
L^{\mu\nu}=\int\frac{d^{4}k}{\left(2\pi\right)^{4}}\bar{u}(p_{1})\frac{1-\gamma_{5}}{2}\gamma_{\mu}\frac{\slashed k_{\nu}+m_{\nu}}{k_{\nu}^{2}-m_{\nu}^{2}}\gamma_{\nu}\frac{1-\gamma_{5}}{2}v(p_{2})e^{i\vec{k}_{\nu}.r}.
\end{equation}
Considering for a moment the two added terms in the propagator, we
see that the two chiral projectors force us to keep only the $m_{\nu}$
term, since:
\begin{eqnarray}
\frac{1-\gamma_{5}}{2}\gamma_{\mu}\slashed k_{\nu}\gamma_{\nu}\frac{1-\gamma_{5}}{2}=\gamma_{\mu}\frac{1+\gamma_{5}}{2}\slashed k_{\nu}\gamma_{\nu}\frac{1-\gamma_{5}}{2}
=\gamma_{\mu}\slashed k_{\nu}\frac{1-\gamma_{5}}{2}\gamma_{\nu}\frac{1-\gamma_{5}}{2}
\\
=\gamma_{\mu}\slashed k_{\nu}\gamma_{\nu}\frac{1+\gamma_{5}}{2}\frac{1-\gamma_{5}}{2}=0.
\end{eqnarray}
So, we find:
\begin{equation}
L^{\mu\nu}=m_{\nu}\int\frac{d^{4}k}{\left(2\pi\right)^{4}}\frac{1}{k_{\nu}^{2}-m_{\nu}^{2}}\bar{u}(p_{1})\gamma_{\mu}\gamma_{\nu}\frac{1-\gamma_{5}}{2}v(p_{2})e^{i\vec{k}_{\nu}.r}.
\end{equation}
Note that if $m_{\nu}=0$ then $L^{\mu\nu}=0$ and the process cannot
go, as expected. Next we can use momentum balance in the denominator
of the propagator to set: 
\begin{equation}
k_{\nu}^{2}-m_{\nu}^{2}=(p_{1i}-p_{n}-p_{1})^{2}-m_{\nu}^{2}.
\end{equation}
In this expression, $p_{1i}$ is the initial momentum of nucleon 1;
$p_{n}$ is the momentum of the intermediate state; and $p_{1}$ is
the momentum of the outgoing electron. Continuing to re-organize:
\begin{eqnarray}
=(E_{1i}-E_{n}-E_{1})^{2}-(\vec{p}_{1i}-\vec{p}_{n}-\vec{p}_{1})^{2}-m_{\nu}^{2}
\\
=(E_{1i}-E_{n}-E_{1})^{2}-\left(\vec{k}_{\nu}^{2}+m_{\nu}^{2}\right)
\\
=(E_{1i}-E_{n}-E_{1})^{2}-\epsilon_{\nu}^{2}.
\end{eqnarray}
Where here $\epsilon_{\nu}$ is the energy of the virtual neutrino.
Continuing to manipulate the leptonic current, we can use the identity:
\begin{equation}
\gamma^{\mu}\gamma^{\nu}=\eta^{\mu\nu}+\frac{1}{2}\sigma^{\mu\nu},\quad\quad(\sigma^{\mu\nu}=\frac{1}{2}[\gamma^{\mu},\gamma^{\nu}]).
\end{equation}
to simplify the leptonic tensor:
\begin{equation}
\\L_{n}^{\mu\nu}=m_{\nu}\int\frac{d^{4}k}{\left(2\pi\right)^{4}}\frac{1}{k_{\nu}^{2}-m_{\nu}^{2}}\bar{u}(p_{1})\left(\eta^{\mu\nu}+\sigma^{\mu\nu}\right)\frac{1-\gamma_{5}}{2}v(p_{2})e^{i\vec{k}_{\nu}.r}.
\end{equation}
For super-allowed transitions, $0^{+}\rightarrow0^{+}$ there is no
``magnetic'' contribution and so we can drop the term proportional
to $\sigma^{\mu\nu}$. This leaves us with:
\begin{equation}
L_{n}^{\mu\nu}=m_{\nu}\eta^{\mu\nu}\int\frac{d^{4}k}{\left(2\pi\right)^{4}}\frac{1}{(E_{1i}-E_{n}-E_{1})^{2}-\epsilon_{\nu}^{2}}\bar{u}(p_{1})\frac{1-\gamma_{5}}{2}v(p_{2})e^{i\vec{k}_{\nu}.r}.
\end{equation}
To do the sum over intermediate states $|n\rangle$ it is typical
to invoke an approximation called the ``closure'' approximation.
This is believed to work well for neutrinoless double beta decay rates
but not so well for two-neutrino double beta decays. The basis of
the approximation is that if it is reasonable to consider all the
intermediate states have approximately the ``mean'' intermediate
state energy $E_{n}\sim\langle E\rangle$ then the weighted sum that
appears in the decay rate simplifies considerably:
\begin{equation}
\sum_{n}\frac{\langle N_{f}|J_{1}|n\rangle\langle n|J_{2}|N_{i}\rangle}{(E_{i}-E_{n}-E_{\eta}-E_{\nu})}\rightarrow\frac{\langle N_{f}|J_{1}\left(\sum_{n}|n\rangle\langle n|\right)J_{2}|N_{i}\rangle}{(E_{i}-\langle E_{n}\rangle-E_{\eta}-E_{\nu})}.
\end{equation}
Because the sum is over a complete set of states, $\sum_{n}|n\rangle\langle n|=1$
and so:
\begin{equation}
\sum_{n}\frac{\langle N_{f}|J_{1}|n\rangle\langle n|J_{2}|N_{i}\rangle}{(E_{i}-E_{n}-E_{\eta}-E_{\nu})}\rightarrow\frac{\langle N_{f}|J_{1}J_{2}|N_{i}\rangle}{(E_{i}-\langle E_{n}\rangle-E_{\eta}-E_{\nu})}.
\end{equation}
Making this approximation we can now factorize the matrix
element into hadronic and leptonic tensors:
\begin{equation}
M=-2G_{F}^{2}\left|V_{ud}\right|^{2}\sum_{n}L_{\mu\nu}^{n}H_{n}^{\mu\nu}\rightarrow-2G_{F}^{2}\left|V_{ud}\right|^{2}L_{\mu\nu}H^{\mu\nu}
\end{equation}
And so the spin-summed matrix element squared needed to calculate
a decay rate factorizes like:
\begin{equation}
\langle|M|^{2}\rangle=4G_{F}^{4}\left|V_{ud}\right|^{4}\left(\sum_{spins}L_{\sigma\rho}L_{\mu\nu}\right)\left(H^{\sigma\rho}H^{\mu\nu}\right).
\end{equation}
Note that each $H$ here contains two powers of $g_A$, so the total number of powers of $g_A$ in the expression is four.  Following a few familiar contractions and lines of rearrangement,
we can get the spin summed product of $L$'s, as:
\begin{equation}
\sum_{spins}L_{\sigma\rho}L_{\mu\nu}=m_{\nu}^{2}g_{\mu\nu}g_{\sigma\rho}Tr\left[\left(\slashed p_{1}+m_{e}\right)\left(\frac{1+\gamma^{5}}{2}\right)\left(\slashed p_{2}-m_{e}\right)\left(\frac{1-\gamma^{5}}{2}\right)\right].
\end{equation}
In this trace, only the elements with both $p$'s survive:
\begin{eqnarray}
=m_{\nu}^{2}g_{\mu\nu}g_{\rho\sigma}(4g_{\alpha\beta}p_{1\alpha}p_{2\beta})
\\
=4m_{\nu}^{2}g_{\mu\nu}g_{\rho\sigma}p_{1}.p_{2}.
\end{eqnarray}
And this can be used to evaluate the matrix element:
\begin{eqnarray}
\langle|{\cal M}|^{2}\rangle=4G_{F}^{2}\left|V_{ud}\right|^{4}H_{\mu}^{\mu}H_{\rho}^{\rho}m_{\nu}^{2}2p_{1}.p_{2}\left[\frac{1}{4\pi}F(r)\right]^{2}\label{eq:MatrixElt}.
\\
F(r)=\frac{1}{4\pi^{2}}\sum_{\eta=1}^{2}\int d^{3}k_{\nu}\frac{e^{ik_{\nu}r}}{(-\epsilon_{\nu})(E_{i}-\langle E_{n}\rangle-E_{\eta}-E_{\nu})}.
\end{eqnarray}
This is the expression for the matrix element if we imagine there
is only one kind of neutrino participating, with mass $m_{\nu}$. 

\subsection{Multiple massive neutrinos in the three flavor paradigm}

In fact in the process of neutrinoless double beta decay, there are
contributions to the decay amplitude from multiple mass eigenstates.
Neglecting neutrino masses within the propagators since they are much
below the energy scale of the process, but keeping the ones in the
numerators, we find an amplitude contribution from each to the matrix element. 

For each neutrino, there is an amplitude contribution that will involve
two factors of the leptonic mixing matrix (a.k.a the PMNS matrix), and the neutrino mass (collecting
into X everything that is not $m_{\nu i}$ of line \ref{eq:MatrixElt}):
\begin{equation}
M_{i}=X\left(U_{ei}\right)^{2}m_{\nu i}.
\end{equation}
And so the total matrix element will be the sum of these contributions,
which ultimately gets squared in Fermi's Golden Rule for the decay
rate:
\begin{eqnarray}
\left|{\cal M}\right|^{2}=\left|X\sum_{i}(U_{ei})^{2}m_{\nu i}\right|^{2}=\left|X\right|^{2}m_{\beta\beta}^{2}
\\
=4G_{F}^{2}\left|V_{ud}\right|^{4}H_{\mu}^{\mu}H_{\rho}^{\rho}2p_{1}.p_{2}\left[\frac{1}{4\pi}F(r)\right]^{2}m_{\beta\beta}^{2}\label{eq:Rate}.
\end{eqnarray}
Above we have introduced the important effective parameter $m_{\beta\beta}$:
\begin{equation}
m_{\beta\beta}=\sum_{i}(U_{ei})^{2}m_{\nu i}.
\end{equation}
Accounting for the matrix element and the final state phase space,
the full decay rate takes the form:
\begin{eqnarray}
\frac{d\Gamma}{d\cos\theta\,dE_{1}}=\left\{ \frac{G_{F}^{4}\left|V_{ud}\right|^{4}}{16\pi^{5}}E_{1}E_{2}|\vec{p}_{1}||\vec{p}_{2}|\left(1-\frac{\vec{p_{1}}.\vec{p_{2}}}{E_{1}E_{2}}\right)\right\} \times
\\
\left\{ H_{\mu}^{\mu}H_{\rho}^{\rho}\left[\frac{1}{4\pi}F(r)\right]^{2}\right\} \left\{ m_{\beta\beta}^{2}\right\}.
\end{eqnarray}
The curly bracketed pieces are called respectively the ``Phase Space
Factor'' (G), the ``Nuclear Matrix Element'' ($\left\Vert M\right\Vert $),
and the ``Effective Majorana Mass'' ($m_{\beta\beta}^{2}$). This
is often written in compact form:
\begin{equation}
\Gamma=G\left\Vert M\right\Vert {}^{2}m_{\beta\beta}^{2}.\label{eq:RateOfBB}
\end{equation}
and for what follows we will absorb the couplings and CKM elements
into $G$:
\begin{equation}
G=G_{F}^{4}\left|V_{ud}\right|^{4}\tilde{G}.
\end{equation}

\subsection{The phase space factor}
In the simplest version of the calculation of $\tilde{G}$, the kinematic
part of $G$, we find:
\begin{equation}
\tilde{G}=\int\frac{1}{16\pi^{5}}dE_{1}d\cos\theta\,E_{1}E_{2}|\vec{p}_{1}||\vec{p}_{2}|\left(1-\frac{\vec{p_{1}}.\vec{p_{2}}}{E_{1}E_{2}}\right).
\end{equation}
This is an integral we can evaluate. First, we explicitly include
the opening angle:
\begin{equation}
=\int\frac{1}{16\pi^{5}}dE_{1}d\cos\theta\,E_{1}E_{2}|\vec{p}_{1}||\vec{p}_{2}|\left(1-\frac{p_{1}p_{2}}{E_{1}E_{2}}\cos\theta\right).
\end{equation}
And note that the integral $\int_{-1}^{1}d\cos\theta\,\cos\theta=0$,
so only the left term survives the angular integration:
\begin{equation}
=\frac{1}{8\pi^{5}}\int_{m_{e}}^{m_{e}+T_{0}}dE_{1}E_{1}E_{2}|\vec{p}_{1}||\vec{p}_{2}|,
\end{equation}
Now using $E_{2}=T_{0}+2m_{e}-E_{1}$, $p_{1}=\sqrt{E_{1}^{2}+m_{e}^{2}}$
and $p_{2}=\sqrt{E_{2}^{2}+m_{e}^{2}}$ we find:
\begin{equation}
=\frac{1}{8\pi^{5}}\int_{m_{e}}^{m_{e}+T_{0}}dE_{1}E_{1}\left(T_{0}+2m_{e}-E_{1}\right)\sqrt{E_{1}-m_{e}^{2}}\sqrt{\left(T_{0}+2m_{e}-E_{1}\right)^{2}-m_{e}^{2}},
\end{equation}
This integral can be done either with blood, sweat and tears, or with
Mathematica, with the result that:
\begin{equation}
=\frac{m_{e}^{5}}{8\pi^{5}}\left(\frac{t_{0}^{5}}{30}+\frac{t_{0}^{4}}{3}+\frac{4t_{0}^{3}}{3}+2t_{0}^{2}+t_{0}\right)\quad\quad t_{0}=\frac{T_0}{m_{e}},\label{eq:PhaseSpaceFactor}
\end{equation}
Where $T_0$ is the total available kinetic energy, labelled in table~\ref{fig:Pairing} as Q$_{\beta\beta}$. Because $T_0>m_e$, the decay rate scales with the fifth power of the available energy, to leading order, all other things being equal.  This is one reason to favor decay isotopes with higher Q-values for experimental study. A second and more pragmatic reason for favoring higher Q$_{\beta\beta}$ is that the higher the Q-value, the more likely the $0\nu\beta\beta$ is to be above the lines from dominant radiogenic backgrounds.

Eq.~\ref{eq:PhaseSpaceFactor} is not quite the end of the story for phase space factors since in the full
calculation we must also include the Fermi function $F(E,Z)$ to account
for the Coulomb attraction of the electron leaving the nucleus~\cite{konopinski1966theory}. This
correction accounts for the fact that the wave function of an electron
leaving as a plane wave is distorted by Coulomb attraction at the
origin. The corrected expression for $\tilde{G}$ is:
\begin{equation}
\tilde{G}=\int\frac{1}{16\pi^{5}}F(E_{1},Z)F(E_{2},Z)dE_{1}d\cos\theta\,E_{1}E_{2}p_{1}p_{2}\left(1-\frac{\vec{p_{1}}.\vec{p_{2}}}{E_{1}E_{2}}\right)
\end{equation}
These coulomb effects must include relativistic corrections and  shielding from the atomic electrons, which makes the integral rather
more complex.  It can be evaluated using numerical methods~\cite{kotila2012phase,mirea2015values}. 

\subsection{The nuclear matrix element}

Calculation of the nuclear matrix elements is far more difficult.
Many different techniques exist, requiring vast computation and with
tracts of supporting literature. The present status of the field is
that the various methods agree on their predictions to within a factor
of 2-3. This is improving as more advanced ab-initio methods reach
maturity, and the computing power to evaluate them becomes more freely
available. Recent reviews of the subject give an excellent coverage
of the various methods in common use~\cite{engel2017status,ejiri2020neutrino}.  The nuclear matrix elements are generally considered the largest of the theoretical uncertainties in predictions of double beta decay rates. 

It is notable also that the effects of $g_{A}$ are buried inside
these objects, to varying degrees, depending on the computational
method chosen. While the uncertainty on $g_{A}$ is sometimes considered
as factorized into a separate question from the uncertainty on calculation
of the matrix element itself, the two cannot be straightforwardly
decoupled. There is evidence from two-neutrino double beta decays
that $g_{A}$ as measured in two neutrino decays may be different
by a large factor, relative to neutron decay, and this introduces
a comparable degree of uncertainty into the predicted rate of $0\nu\beta\beta$
to other aspects of the matrix element calculation. This situation
is evolving, and for a relatively modern discussion see Ref~\cite{suhonen_jouni_2018_1286917}.

\begin{figure}
\includegraphics[width=0.49\columnwidth]{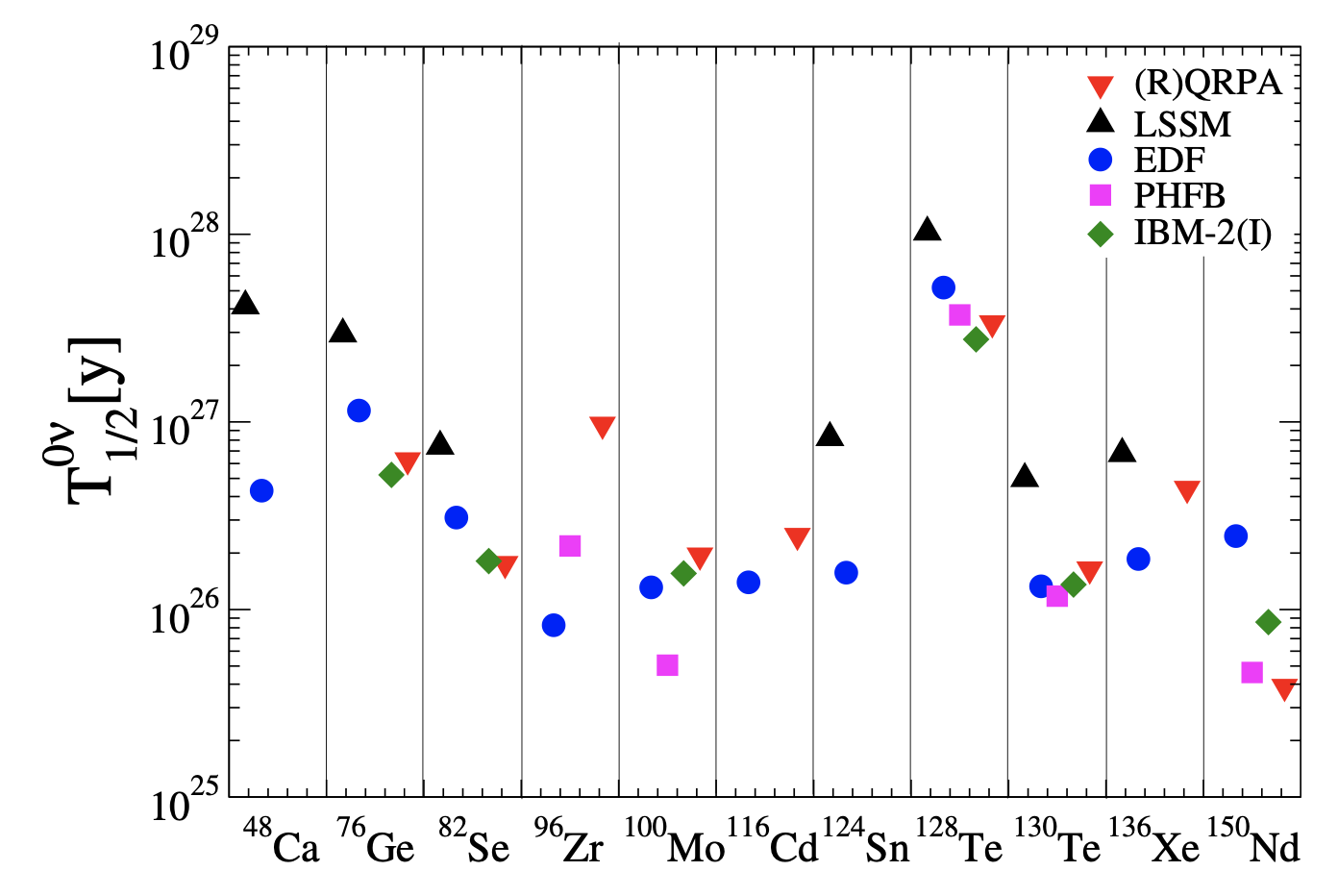}\includegraphics[width=0.5\columnwidth]{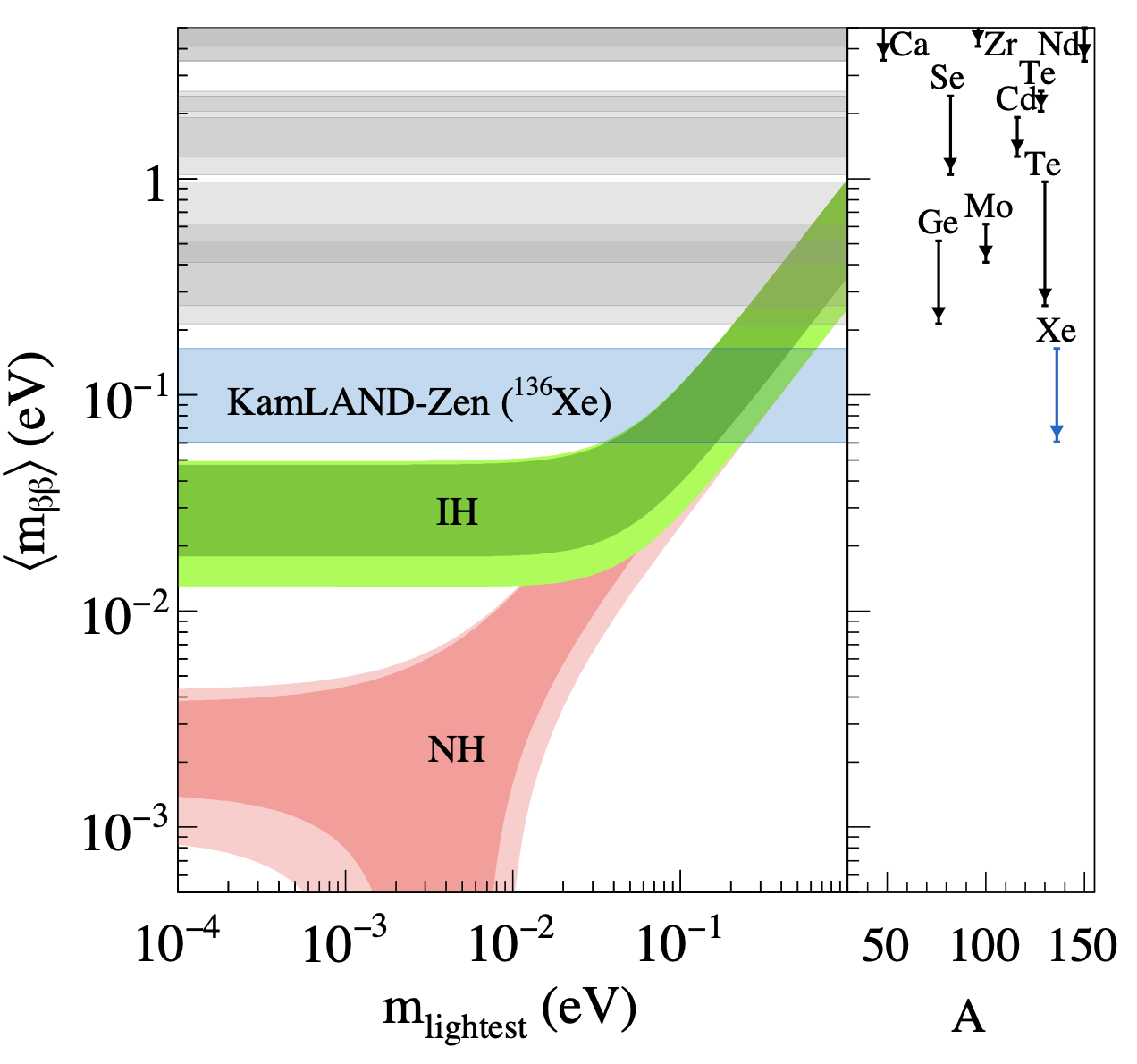}

\caption{Left: Predictions for the half life of 0nubb in a given reference model
with different calculated nuclear matrix elements. Right: Dependence
of the parameter $m_{\beta\beta}$ on the presently unknown lightest
neutrino mass, with bands showing allowed regions given what is known
about mixing parameters. \label{BandPlot}}
\end{figure}

\subsection{The effective Majorana mass}The effective mass that shows up in double beta decay rate of Eq.
\ref{eq:Rate} is a weighted sum of the three neutrinos masses, each
contributing proportionally to their probabilistic weight within the
electron neutrino flavor state:
\begin{equation}
m_{\beta\beta}=\sum_{i}(U_{ei})^{2}m_{i}\label{eq:MassSum}.
\end{equation}
We can express the matrix elements $U_{ei}$ in terms of the mixing
angles and phases that traditionally parameterize the PMNS matrix:
\begin{equation}
U_{\alpha i}=\left[\begin{array}{ccc}
1 & 0 & 0\\
0 & c_{23} & s_{23}\\
0 & -s_{23} & c_{23}
\end{array}\right]\left[\begin{array}{ccc}
c_{13} & 0 & s_{13}e^{-i\delta}\\
0 & 1 & 0\\
-s_{13}e_{-i\delta} & 0 & c_{13}
\end{array}\right]\left[\begin{array}{ccc}
c_{12} & s_{12} & 0\\
-s_{12} & c_{12} & 0\\
0 & 0 & 1
\end{array}\right]\left[\begin{array}{ccc}
e^{i\lambda_{a}} & 0 & 0\\
0 & e^{i\lambda_{b}} & 0\\
0 & 0 & 1
\end{array}\right]
\end{equation}
Multiplying out all the terms in of Eq.~\ref{eq:MassSum} then yields:
\begin{equation}
m_{\beta\beta}=c_{12}^{2}c_{13}^{2}e^{2i\lambda_{a}}m_{1}+c_{13}^{2}s_{12}^{2}e^{2i\lambda_{b}}m_{2}+s_{13}^{2}m_{3}.
\end{equation}
Let us briefly review what we know about the quantities in this equation~\cite{smoot2020review}.  Regarding the neutrino masses, all we know today are the their squared
differences, accessed through oscillations. We do not know the absolute
mass scale (the lightest $m_{i}$) or whether the observed bigger
splitting is between the heaviest two or lightest two neutrinos (the
``mass ordering'' or ``mass hierarchy''). We do know all of the
mixing angles, with reasonable precision, from studies of neutrino
oscillations between various flavors and on various baselines. We
might now know something slightly more than nothing about $\delta_{CP}$ - if we do then this
is recent news~\cite{abe2018search}. We certainly know
nothing about the Majorana phases $\lambda_{a}$ and $\lambda_{b}$
since they do not feature in oscillation probabilities, and we have little hope
of learning about them, short of observing neutrinoless double beta
decay.

In terms of these known and unknown parameters $m_{\beta\beta}$ can
be expressed as:
\begin{equation}
m_{\beta\beta}=c_{12}^{2}c_{13}^{2}m_{1}e^{2i\lambda_{a}}+s_{12}^{2}c_{13}^{2}e^{2i\lambda_{b}}\sqrt{m_{1}^{2}+\Delta m_{12}^{2}}+s_{13}^{2}\sqrt{m_{1}^{2}\pm|\Delta m_{23}^{2}|}\label{eq:mbb}.
\end{equation}
The $\pm$ under the square root of Eq. \ref{eq:mbb} reflects that
at the present time we know the absolute scale of $\Delta m_{23}^{2}$
(from atmospheric and accelerator neutrino experiments) but we do
not know its sign (the ``mass ordering'', or ``mass heirachy'').
On the other hand, $\Delta m_{12}^{2}$ is known from solar neutrino
oscillation experiments where the MSW effect would drive oscillations
differently depending on the relevant ordering, so we do know both
its value and sign. 

Given the freedom to choose all the unknown parameters $\lambda_{a},\lambda_{b},m_{1}$
in Eq.\ref{eq:mbb}, as well as make one discrete choice of the sign
of $\Delta m_{23}^{2}$, we find two swathes of allowed decay
rates. These bands are commonly represented on what has become colloquially
known as ``lobster plot'' of Fig. \ref{BandPlot}, right. Here the
allowed values for the parameter $m_{\beta\beta}$ featuring in the
decay rate (or equivalently the lifetime) is shown with its allowed
values plotted against the lightest neutrino mass. The lifetime of
neutrinoless double beta decay is proportional to $|m_{\beta\beta}|^{2}$. 

\section{Mechanisms of neutrinoless double beta decay and the Schechter Valle
theorem}

We must mention an important point at this juncture. The argument
presented so far went as follows: if the known neutrinos are Majorana
particles, they will induce neutrinoless double beta decay. This decay
will occur at a rate that depends in a well-defined way on $m_{\beta\beta}$,
shown in the right plot of Fig. \ref{BandPlot}. This is true so long
as there are three light Majorana neutrinos and no other lepton-number-violating
physics that contributes to the decay. Both of these assumptions deserve
scrutiny.

\begin{figure}
\begin{centering}
\includegraphics[width=0.8\columnwidth]{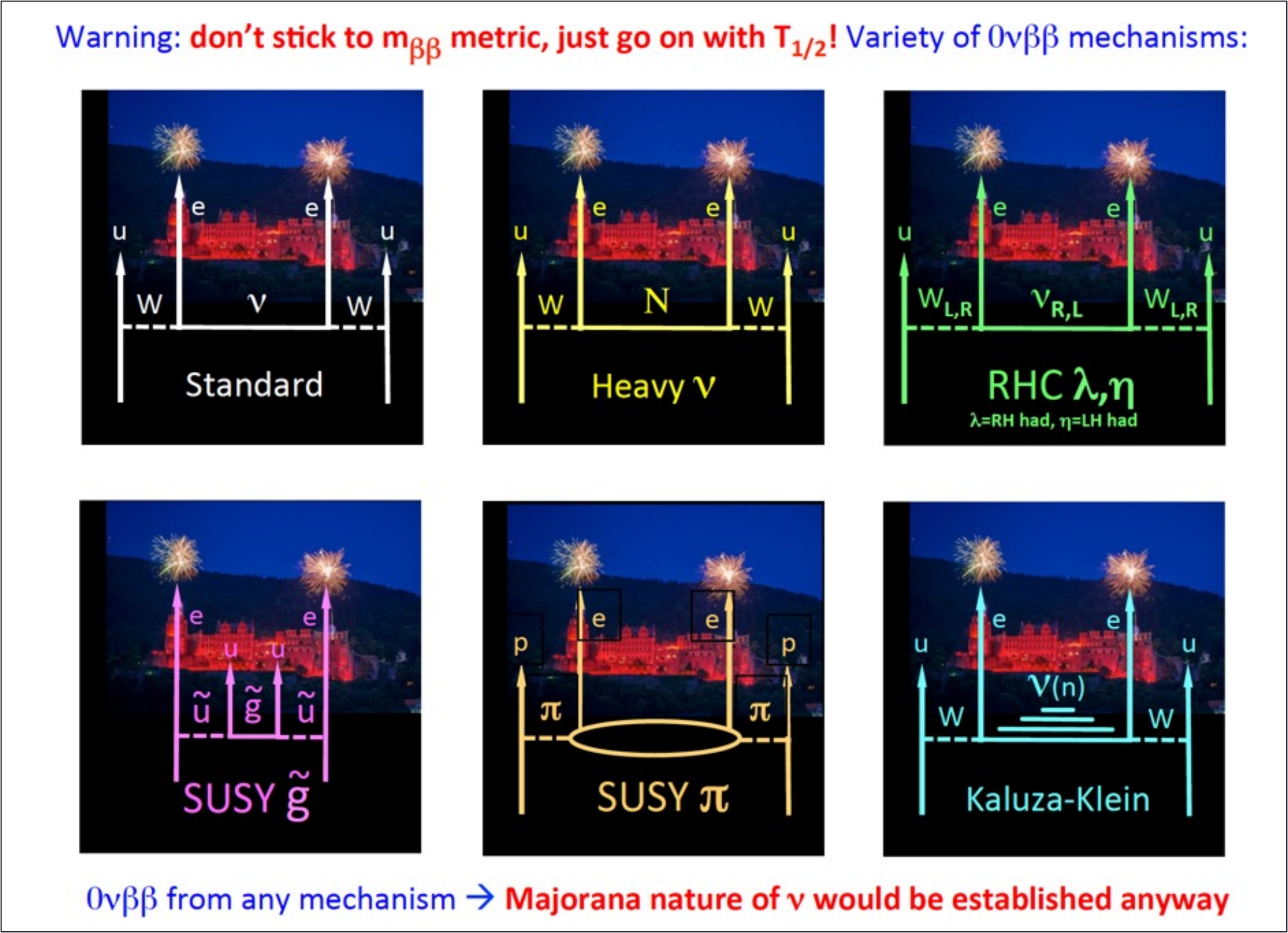}
\par\end{centering}
\caption{Various mechanisms of $0\nu\beta\beta$, borrowed from ~\cite{lisi_eligio_2018_1286745}. \label{fig:Various-mechanisms-of}}
\end{figure}

First, some short baseline experiments have generated anomalies that
may be interpreted as evidence of new, heavier neutrino mass states.
The corresponding flavor states must be sterile due to constraints
from the invisible width of the Z boson~\cite{acciarri1998determination}. Evidence for sterile neutrinos~\cite{abazajian2012light}
is presently inconclusive, and there are large tensions between positive
and negative observations. Should light sterile neutrinos exist and
be Majorana particles they would add an additional mass state to the
sum in Eq. \ref{eq:MassSum}, invalidating Eq. \ref{eq:mbb}. In the
presence of sterile neutrinos, the decay rate may be either much larger
or much smaller than the predicted three-neutrino rate in either ordering~\cite{giunti2015predictions},
for a given set of $\lambda_{a},\lambda_{b},m_{1}$, depending on
the value of an additional Majorana phase. Existence of a sterile
neutrino would move (and widen) the goalposts in neutrinoless double beta decay
dramatically.

Furthermore, while light Majorana neutrino exchange may induce neutrinoless
double beta decay, it does not necessarily have to be the only mechanism
driving the process. There are a wide variety of lepton number violating
sources of new physics that may occur within a nucleus and drive $0\nu\beta\beta$.
As a fairly general statement, if you have a theory with high scale
lepton number violation that introduces effective operators into the
low energy Lagrangian at dimension 7, 9, 11, etc~\cite{cirigliano2017neutrinoless,cirigliano2018neutrinoless}, there is a good chance
it will be able to drive neutrinoless double beta decay. If these
more complex mechanisms are responsible for $0\nu\beta\beta$, the
relationship between the decay rate and neutrino mixing parameters and
masses will, of course, not follow Eq.~\ref{eq:RateOfBB}, and could be much larger. This is the case
no matter which mass ordering or lightest neutrino mass nature has
chosen.

It might seem far-fetched to invoke exotic new physics scenarios as
a cause for optimism, to bump up the predicted rate of $0\nu\beta\beta$
above what is suggested by the standard mechanism.  It is an especially appealing
thing for experimentalists to imagine, especially if we happen to live in a normal-mass-ordered
scenario where the lifetimes that must be probed to find the standard
mechanism are truly formidable.  But is it just wishful thinking, or is there really reason to be hopeful?  This is, of course, a question without a truly rigorous answer - but is worth noting that
the conventional seesaw mechanism would set the energy scale of neutrino
mass generating physics relative to the electroweak scale $\Lambda_{EW}$
at something like $\Lambda_{N}=\Lambda_{EW}^{2}/m_{\nu}\sim10^{23}\mathrm{eV}$.
This is far above the energy scale of most new physics scenarios being
sought at (though admittedly not yet found at) the Large Hadron Collider, and many of those scenarios are motivated by the need for
TeV-scale physics to resolve the Heirachy problem that leads to quadratic
corrections to the Higgs boson mass. Based on purely dimensional arguments,
lepton number violation from such lower-scale processes would drive
a faster rate of $0\nu\beta\beta$ than the conventional light Majorana
neutrino exchange mechanism. Some examples are shown in Fig. \ref{fig:Various-mechanisms-of}.

This discussion of the various possible mechanisms may cause one to
wonder: if we see $0\nu\beta\beta$, what did we actually learn? If
it could be caused by any new physics whatsoever, would we know anything
other than ``neutrinoless double beta decay happens'' if we were to see it?  Will we even know the 
neutrino is a Majorana fermion?  

\begin{figure}
\begin{centering}
\includegraphics[width=0.6\columnwidth]{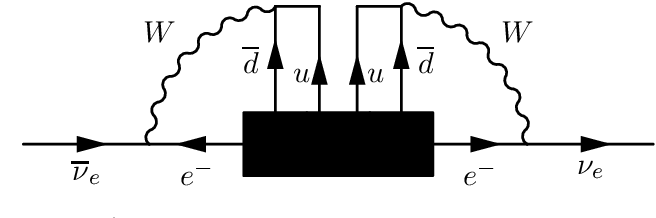}
\par\end{centering}
\caption{The Schechter Valle Feynman diagram that leads to a Majorana mass, given any new physics that generates $0\nu\beta\beta$\label{fig:Schechter}}
\end{figure}

To begin with, we would obviously know right away that lepton number is violated, since
we would have observed a lepton number violating process. But it turns
out we would know more than this.  It turns out that if double beta decay occurs, it is necessary that the neutrino is a Majorana fermion, even
if the dominant mechanism causing $0\nu\beta\beta$ were not light Majorana
neutrino exchange mechanism we have discussed. The connection is made by the Schechter Valle
theorem~\cite{schechter1982neutrinoless}, which says that given any possible source of lepton number
violating physics that causes $0\nu\beta\beta$, one can draw a Feynman
diagram (Fig.~\ref{fig:Schechter}) enclosing that new physics as an internal component, whose
outcome is the generation of a Majorana neutrino mass. Because of
this theorem the logic is bidirectional: existence of Majorana neutrinos
implies $0\nu\beta\beta$, if by no other mechanism than at least
by light Majorana neutrino exchange; and the existence of $0\nu\beta\beta$
implies Majorana neutrinos masses, generated at least by via Schechter Valle
diagram, if nothing else.  Observation of $0\nu\beta\beta$ means neutrinos are Majorana; period.

\section{Backgrounds and sensitivities to neutrinoless double beta decay}

Backgrounds to $0\nu\beta\beta$ only partially derive from the two
neutrino process. In experiments that meet the energy resolution criterion
$\mathrm{FWHM}\leq2\%$\footnote{Note that some experiments report standard deviations rather than
FWHM, and the two can be related by $\sigma=\mathrm{FWHM}/2.35$.}, backgrounds will generally be dominated by radiogenic processes.
These primarily involve gamma rays from the uranium and thorium chains.
There is a lot we could say about these backgrounds, about how they
can be simulated, minimized, rejected and generally mitigated. For
our present purposes, what we need to know is: to have a truly sensitive
experiment one must drive these backgrounds down, through either selection
of clean materials, use of powerful new technologies, or advanced
data analysis methods, to below $B\sim0.1$ counts per ton per year
in the energy region-of-interest, and that is very hard to do. 

\begin{figure}[t]
\begin{centering}
\includegraphics[width=0.7\columnwidth]{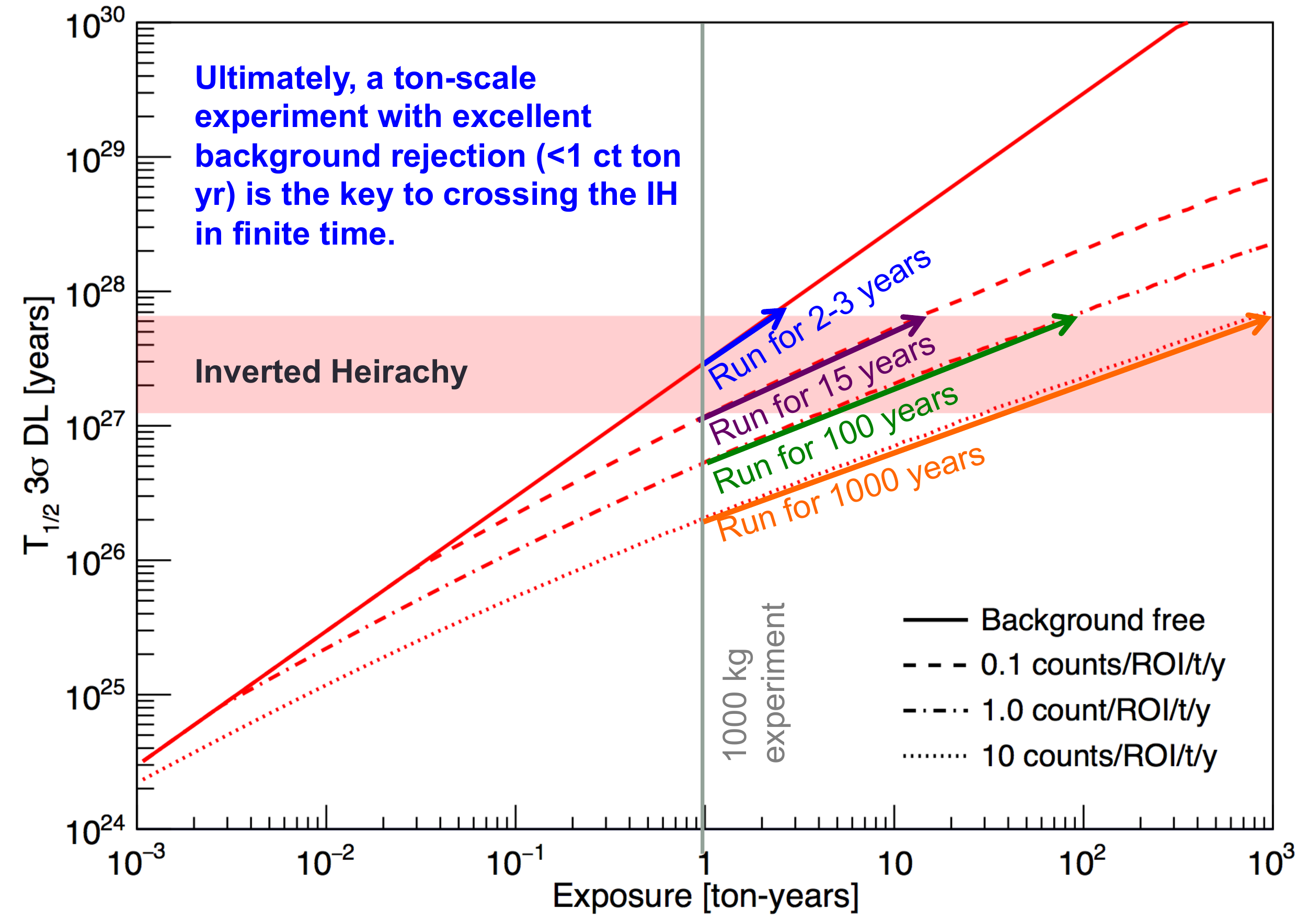}
\par\end{centering}
\caption{Sensitivity vs background index for neutrinoless double beta
decay, adapted from~\cite{agostini2017discovery}.
\label{fig:Left:-sensitivity-vs}}
\end{figure}

The reason this is important is shown in Fig.~\ref{fig:Left:-sensitivity-vs},
top. This vertical axis of this figure shows the growth in $3\sigma$
discovery potential, defined as the half-life which one would expect
to make a $3\sigma$ discovery of in 50\% of experimental searches, were the signal real. We can only make this statement statistically since $0\nu\beta\beta$
is a random process and any given nucleus has some probability for decaying and some probability for not decaying while
we are looking, no matter how long we wait. The horizontal axis shows exposure, defined as mass
times run-time $\epsilon=Mt$: having more of either means more expected
signal. 

The sensitivities for low exposures are all proportional to
$T_{1/2}$. In this low-exposure regime, the number of background
events expected is much less than one; thus observation of a single
event would be a high-significance a discovery, since it must be signal.
Doubling the exposure in this regime effectively doubles the half-life
that corresponds to a 50\% chance of an event in this time window,
explaining the proportionality. Since $T_{1/2}$ is inversely proportional
to $m_{\beta\beta}^{2}$, sensitivity of neutrinoless double beta
decay experiments initially grows with time as $T_{1/2}\propto\epsilon$, or $m_{\beta\beta}\propto\epsilon^{-1/2}$. 

At larger exposures, the lines all transition to being to proportional
to $\sqrt{T_{1/2}}$. In this regime the experiment has been running
long enough that some background events are expected, and the question
becomes one of finding a statistical excess: what is the probability of a given rate of signal events $S$ over particular rate of background. In the high-exposure
limit, the number of expected background events $B$ increases proportionally
to $\epsilon$, as does the number of expected signal events, so sensitivity
grows like $T_{1/2}\propto\sqrt{\epsilon}$, or $m_{\beta\beta}\propto\epsilon^{-1/4}$.
When experiments reach this regime their progress in sensitivity is
thus exceedingly slow, and returns diminish fast. The determining
factor in how soon a given experiment will reach this turning point
is the level of background. This is what distinguishes the four curves
shown in Fig. \ref{fig:Left:-sensitivity-vs}.

\begin{figure}[t]
\begin{centering}
\includegraphics[width=0.8\columnwidth]{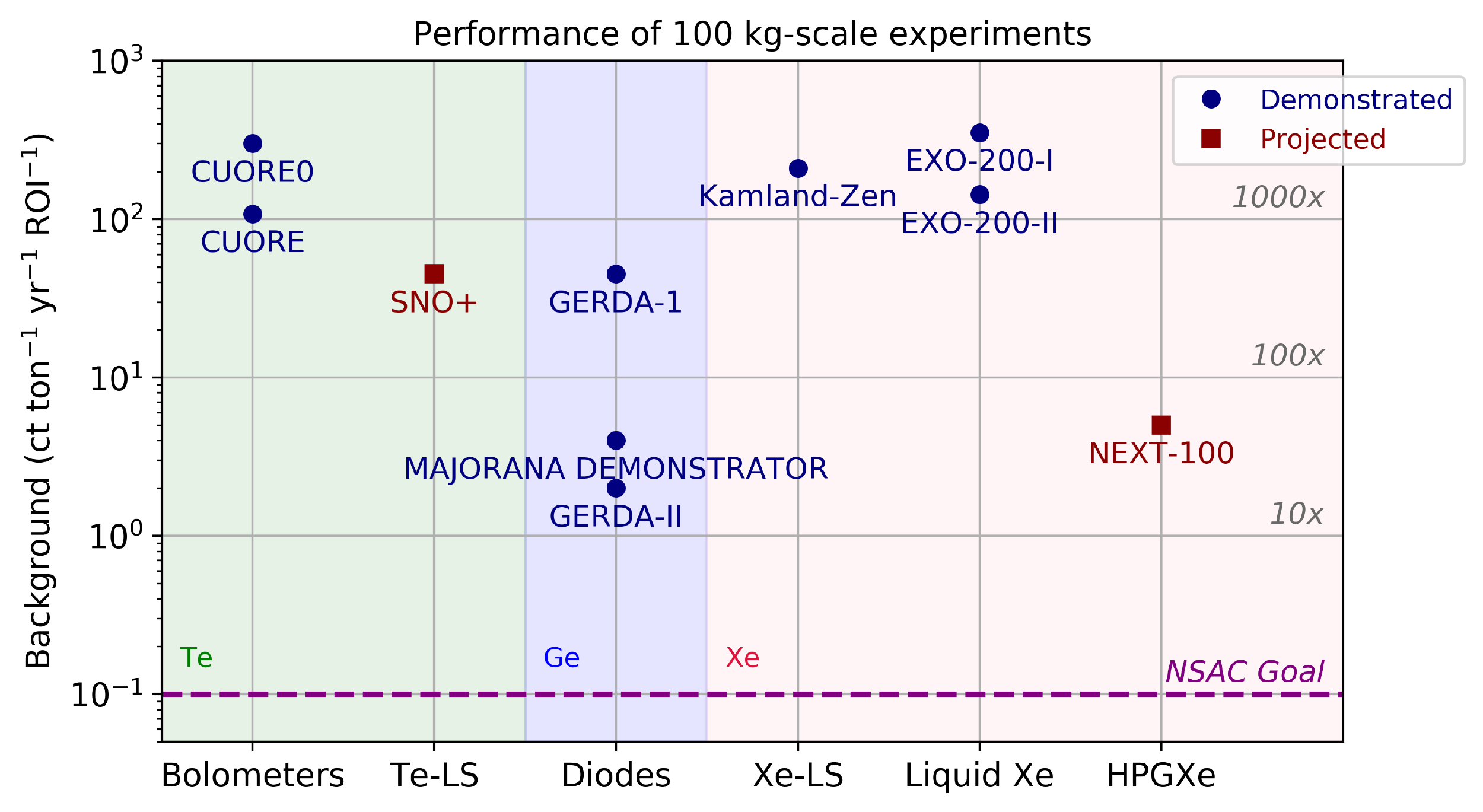}
\par\end{centering}
\caption{Approximate background indices in present (100kg-scale) programs.
\label{fig:BGIndices}}
\end{figure}

We see from this figure that to cross the inverted ordering parameter
space in reasonable time relies on riding the background-free line
as far as possible, with background indices of at most B\textasciitilde 0.1,
for a ton-scale experiment. We emphasize again though, that this inverted-ordering
band is only a visual guide, and discovery of $0\nu\beta\beta$ is
in principle possible given either ordering at any value of the half-life
beyond existing limits ($10^{26}\mathrm{yr}$ in $^{136}Xe$, for example). Once the rate of background becomes
comparable to the expected signal, further progress is
substantially more difficult. Fig.~\ref{fig:BGIndices} shows
the demonstrated or projected background rate of existing 100kg scale
experiments. It is clear that all experiments to date have been in
the strongly background limited regime, when extrapolated to ton-scale
technologies at their existing levels of background.

The world's leading experimental limit on $0\nu\beta\beta$ is from
the Kamland-Zen experiment~\cite{gando2016search}, which searches for the decay in $^{136}$Xe-doped liquid scintillator
and has set a lifetime limit of $\tau\geq1.07\times10^{26}\mathrm{yr}$
at 90\% confidence level. This corresponds to a 90\% CL upper limit
of $m_{\beta\beta}\sim$61-165 meV with the range depending primarily
on the nuclear matrix elements assumed. Other, somewhat less strong
limits have been obtained using other isotopes including $^{76}$Ge
and $^{130}$Te. Ongoing R\&D now aims to realize lower background
technologies for ton- to multi-ton scale experiments.

Leading the field in terms of background index at the present time
are germanium diodes with $b\sim2$ counts per ton per year in the energy
region of interest, pioneered by the GERDA~\cite{agostini2020final} and Majorana Demonstrator~\cite{aalseth2018search}
collaborations. The ton-scale phase of these programs is being pursued
as a unified international collaboration, via the LEGEND program~\cite{abgrall2017large}.
Improved background indices beyond the presently demonstrated $b\sim200$
in liquid xenon from the EXO-200~\cite{albert2018search} collaboration are being pursued via
the use of dramatic self-shielding of liquid xenon, via the nEXO~\cite{albert2018sensitivity} collaboration,
which aims to mount a 5-ton scale liquid xenon time projection chamber,
with the goal of improving on the EXO-200 background index by a factor
of over 1000. The CUPID~\cite{azzolini2018first} collaboration is pursuing a program of development
of scintillating bolometers to remove the major background sources
from radioactive decays of isotopes plated onto the crystal surfaces in the CUORE tellurium
bolometer experiment. And the NEXT collaboration~\cite{lopez2017sensitivity} aims to realize a
low background high pressure xenon gas time projection chamber at
the ton-scale~\cite{adams2021sensitivity}, using measurements of both energy and event topology
(tracking of two electrons in double beta decay events rather than
one from radiogenic backgrounds) to achieve unprecedentedly low background
indices with the isotope $^{136}$Xe. New technologies, including
methods of identifying the $^{136}$Ba ion emitted in the double beta
decay of $^{136}$Xe are also under development~\cite{nEXOSingle,mcdonald2018demonstration}, which aim for realization
of ultra-low-background, or potentially even zero-background technologies
at the ton- to multi-ton scale. Fundamental advances such as these
are likely to be required in order to penetrate half-lives of the
other $10^{28}$ years, as suggested, for example, by the light Majorana neutrino
exchange model given a normal neutrino mass ordering.

\section{Conclusions}

The question of the nature of the neutrino mass is one of profound
scientific importance. A discovery of a Majorana nature to the neutrino
would confirm the standard model as a low energy effective theory;
demonstrate that there are objects in the universe that are neither
matter or antimatter but some strange hybrid of the two; illuminate
a new mass-generating mechanism for fundamental particles beyond the
Higgs mechanism alone; confirm the existence of lepton number violation
in nature; and provide support for leptogenesis as the mechanism that
generated the observed matter-antimatter asymmetry of the Universe. 

Tests of the Majorana nature of the neutrino rely
on searching for the hallmark of Majorana neutrinos, lepton number
violation, in processes where the neutrinos are intermediate particles
that do not need to be directly observed. Achieving sensitivity at
high energy accelerator experiments is implausible given what is known
about the mass of the neutrino and the chiral properties of the weak
interaction. However, an ultimate short-baseline experiment is possible:
neutrinoless double beta decay. In this process, exchange of a neutrino
between two nucleons in a nucleus leads to production of two electrons
and no neutrinos in the final state. If observed, this process would
demonstrate the neutrino to be a Majorana particle.

Observation of $0\nu\beta\beta$ is formidably hard, because the process
is expected to be extremely slow. However, a range of low-background
technologies have been developed that have pursued searches $0\nu\beta\beta$
with sensitivities that have reached beyond $10^{26}$ years in half-life.
Now, advanced methodologies and ton- to multi-ton scale experiments
are being pursued to push this sensitivity still further, with proposed
programs aiming to reach near to $10^{28}$ years in half-life. These
experiments are extremely challenging, but the scientific payoff of
a discovery would be profound. The quest to discover $0\nu\beta\beta$
continues to motivate an international community of physicists to
develop advanced technologies, materials and analysis methods, with
the goal of achieving a scientific discovery with profound implications for particle physics, nuclear physics and
cosmology.

\section*{Acknowledgements}

    BJPJ thanks Manuel Tiscareno for proof-reading and type-setting assistance, and UTA students Matthew Molewski, Karen Navarro, Ivana Moya, Jackie Baeza Rubio, Tyler Workman, Logan Norman and Ben Smithers for astute and important comments on the manuscript.  We also thank the TASI conference organizers for the invitation to give these lectures, and the flexibility to deliver the proceedings exceedingly late due to the difficulties of writing during the year of COVID-19.  BJPJ's work on the NEXT program is supported by the Department of Energy under Early Career Award number DE-SC0019054.
    
\bibliographystyle{JHEP}
\bibliography{biblio}

    \end{document}